\documentclass[aps,prd,twocolumn,showpacs,superscriptaddress,groupedaddress]{revtex4}  

\usepackage{amsmath}
\usepackage{amsfonts}
\usepackage{bm}
\usepackage{color}
\usepackage{graphicx}
\definecolor{darkgreen}{cmyk}{0.85,0.2,1.00,0.2}

\usepackage{amsbsy}

\usepackage{amsmath,amssymb,subfigure}
\usepackage{graphicx}
\usepackage{epsfig}
\usepackage{color}
\usepackage{ulem}
\usepackage{epstopdf}
\newcommand{\be}{\begin{eqnarray}}

\newcommand{\ee}{\end{eqnarray}}

\newcommand{\D}{\mathcal{D}}

\newcommand{\R}{\mathcal{R}}

\newcommand{\ls}{\mathrel{\raise0.27ex\hbox{$<$}\kern-0.70em \lower0.71ex\hbox{{$\scriptstyle \sim$}}}}

\begin{document}

\title{Alleviating the tension at low $\ell$ through Axion Monodromy}

\author{P.~Daniel Meerburg} 
\affiliation{Department of Astrophysical Sciences, Princeton University, Princeton, NJ 08540 USA. }
\email{meerburg@princeton.edu}
\date{\today}

\begin{abstract} 
There exists some tension on large scales between the Planck {\it data} and the $\Lambda$CDM concordance model of the Universe, which has been amplified by the recently claimed discovery of non-zero tensor to scalar ratio $r$. At the same time,  the current best-fit value of $r$ suggests large field inflation $\Delta \phi_{\rm inf}>M_p$, which requires a UV complete description of inflation. A very promising working example that predicts large tensor modes and can be UV completed is axion monodromy inflation. This realization of inflation naturally produces oscillating features, as consequence of a broken shift symmetry. We analyse a combination of Planck, ACT, SPT, WMAP low $\ell$ polarization and BICEP2 data, and show a long wavelength feature from an periodic potential can alleviate the tension at low multipoles with an improvement $\Delta \chi^2\simeq 2.5-4$ per degree of freedom, depending on the level of foreground subtraction. As with an introduction of running, one expects that any scale dependence should lead to a worsened fit at high multipoles. We show that the logarithmic nature of the axion feature in combination with a tilt $n_s\sim1$ allows the fit to be identical to a no-feature model at the $2$ percent level on scales $100 \leq \ell \leq 3500$, and quite remarkable actually slightly {\it improves}  the fit at scales $\ell >2000$. We also consider possible unremoved dust foregrounds and show that including these hardly changes the best-fit parameters. Corrected for potential foregrounds and fixing the frequency to the best fit value, we find an amplitude of the feature $\delta n_s = 0.095 ^{+0.03}_{-0.05}$, a spectral index $n_s = 1.0^{+0.03}_{-0.04}$, the overall amplitude $\log 10^{10}A_s =3.06 \pm 0.04$ and a phase $\phi = 0.85 ^{+0.9}_{-1.6}$. These parameters suggest an axion decay constant of $f/ M_p \sim \mathcal{O}(.01)$. We discuss how Planck measurements of the TE and EE spectra can further constrain axion monodromy inflation with such a large feature. A measurement of the large scale structure power spectrum is even more promising, as the effect is much bigger since the tensor modes do not affect the large scales. At the same time, a feature could also lead to a lower $\sigma_8$, lifting the tension between CMB and SZ constraints on $\sigma_8$.

\end{abstract}

\maketitle
\section{Introduction}

The recent discovery of B-mode polarization in the sky \cite{BICEP2014} has created enormous excitement in the community. Foremost, if the B-modes are sourced by primordial gravitational waves, the discovery can be considered additional evidence for the inflation paradigm; in the very early Universe, a yet unknown scalar field dominated the energy density in the Universe causing the Universe to expand at least $60$-efolds in a very short time. Secondly, the associated amplitude of the primordial signal is bigger than anticipated from earlier experiments, where the constraint is derived from the CMB temperature power spectrum. In a $\Lambda$CDM + $r$ concordance cosmology, Planck \cite{planckcosmoparms2013} derived that $r_{0.05}<0.135$ at $95 \%$ \footnote{As pointed in \cite{Audren2014} these numbers need to be compared when $n_t=0$, leading to a $1.7\sigma$ discrepancy. }. The BICEP experiments derives a non-zero value of $r_{0.05}=0.2_{-0.05}^{+0.07}$, which is almost $2$ sigma away from this value. It is obvious then, that this adds to the existing tension on large angular scales between the data and the $\Lambda$CDM + $r$ concordance cosmology, as gravitational waves do not only source B-modes but also add to the scalars. Though this part of the spectrum has large cosmic variance, if the large value of $r$ persists, we will  possibly have to reconsider a simple power law inflation. Naturally any power law will have some higher order deviation from scale invariance beyond the tilt. However, for any polynomial power law, the  running will always be slow-roll suppressed compared to the tilt \footnote{Sums of altering sign power laws can do allow for a scale dependence that leads to a different hierarchy \cite{Choudhury2014}.}. Besides, when considering the BICEP evidence for a large $r$, a running does not lead to a significantly better fit on large scales, and at the same time it will lead to a worsened fit on small scales (see e.g.  \cite{Abazajian2014,Hazra2014} for several recent attempts \footnote{In this paper \cite{2014arXiv1405.2784W}, an oscillation is considered and shown to be a good fit, however without an actual data analysis, while in \cite{2014arXiv1403.5277F} natural inflation is considered, but is not fitted for. }). Thirdly, the large tensor to scalar ratio indicates that the field driving inflation has been displaced overall several $M_p$, which can be considered an issue from a model building perspective. It would be interesting if there exists a model that fits the data in 3 ways; it has naturally large tensor modes, it has some sort of running on large scales (and possibly on not yet observed small scales) and deals with super Planckian displacements in field space. In this paper we will show that such a model already exists in the form of axion monodromy inflation and it will turn out that it actually provides a significantly better fit to the data at large scales and even improves the fit on the very small scales. In addition, it makes clear testable prediction for the TE, EE and large scale structure power spectra, which should be measured with running and upcoming experiments. Interestingly, a large feature also lowers the derived value of $\sigma_8$, relieving some of the tension with measurements of $\sigma_8$ using the SZ signal \cite{PlanckSZ2013}. 

This paper is organized as follows. We will briefly give a theoretical background and motivation to consider a large scale (i.e. long wavelength) feature in light of the recent BICEP2 measurements in \S\ref{thbackground}. In \S\ref{Results}, we present our results and show that a long wavelength feature provides an improved fit to the data. We will consider the most pessimistic foreground models derived in \cite{BICEP2014}  in combination with large and small scale CMB experiments. Our findings will be discussed in \S\ref{discussion}, where we will present some simple estimate of how much a large feature could be due to cosmic variance. In addition, we will discuss the possible implications for EE and TE polarization as well the matter power spectrum. We conclude in \S\ref{Conclusion}.

\section{Theoretical Background} \label{thbackground}
Under the assumption that the discovery of non-zero B-modes in the CMB sky \cite{BICEP2014} is caused by the production of  primordial gravitational waves (see \cite{2014arXiv1405.7351F} for a thorough discussion on this topic), together with the COBE normalization $\delta_H\simeq 1.9\times10^{-9}$ we can infer the scale of inflation 
\be
\Lambda \simeq \left( \frac{r}{0.2}\right)^{1/4} 2.4 \; 10^{16}\; {\rm Gev},
\ee 
which implies $\Lambda \sim 10^{16}$ GeV for the current measured $r\simeq 0.2$. In addition, besides a measurement of $V=\Lambda^4$, through $r = 16\epsilon$, and $\epsilon\propto(V'/V)^2$, we also learn about $V'$. As such, the detection of $B$-modes has already allowed one to limit the number of inflationary models, while the simplest model has persisted. However simple, the measurement of large $r$ has implications for inflation beyond the presence of primordial gravitational waves. According to the {\it Lyth bound} \citep{Lyth1999,EastherKinney2006}  
\be
\Delta \phi_{\rm inf} \geq M_p \sqrt{\frac{r}{4\pi}}.
\ee
The bound is valid for slow-roll single-field inflationary models but can be generalized \citep{BaumannGreen2012}. The main idea is that the production of observable gravitational waves can only be accomplished during inflation if the field driving inflation displaces over super Planckian distances, during the time the observable scales exit the horizon. This approximate theoretical bound poses a challenge for model builders, as it requires one to fully formalize a model that is UV complete; the absence of such prescription, would introduce an infinite fine tuning of higher order operators. Although this is only a theoretical complaint, it is considered unsatisfying and to some extend nonpredictive, arguably driving us towards an anthropic point of view; the precise cancelation of higher order terms can only be explained by assuming that enough E-folds of inflation would not have occurred without it and hence, we would have not been here to observe it. 
\begin{figure}[htbp] 
   \centering
   \includegraphics[width=3in]{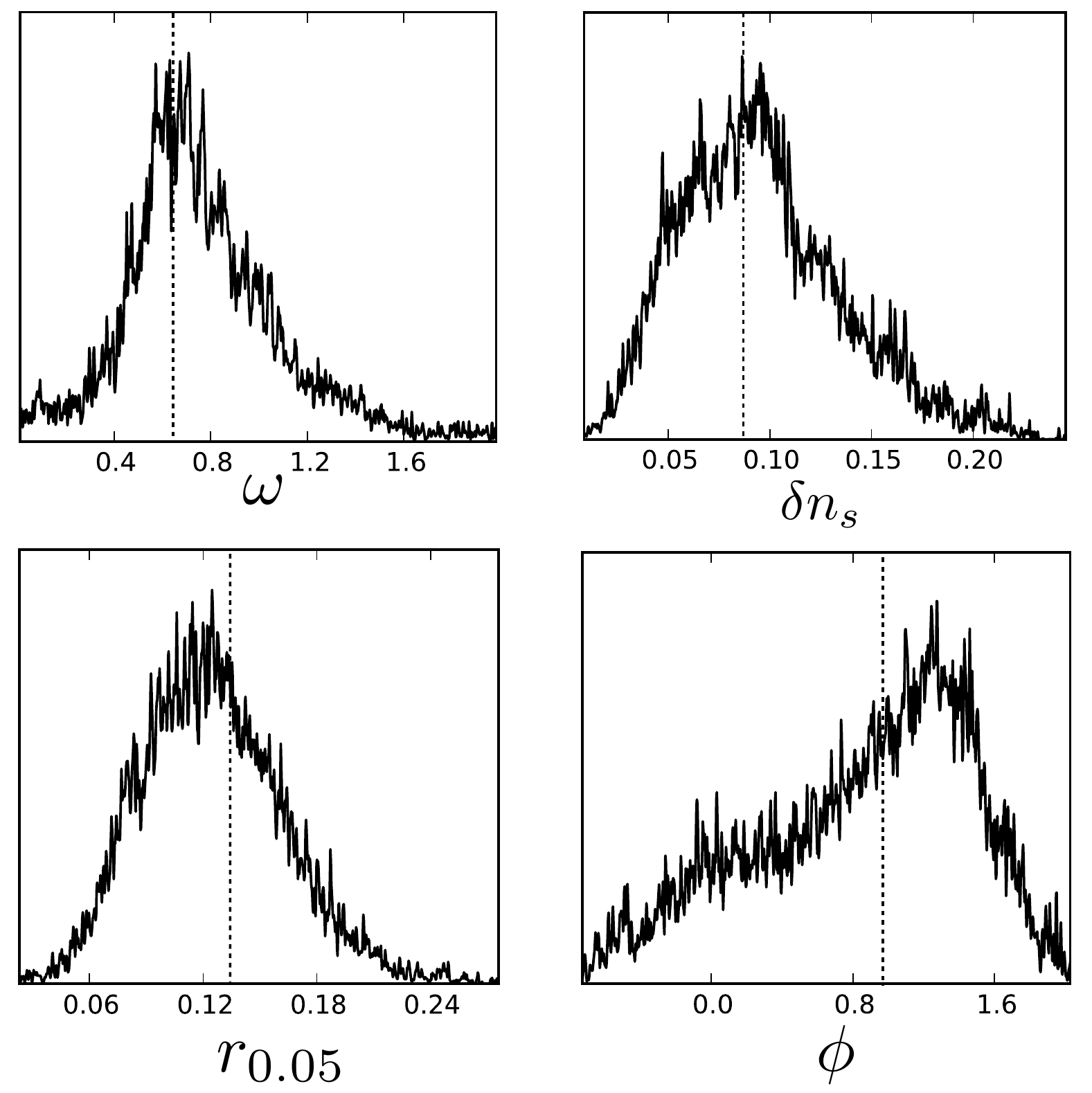} 
   \caption{Marginalized posterior probability distribution for $\omega$, $\delta n_s$, $r_{0.05}$ and the phase $\phi$ derived from Planck+Lensing+WP+high $\ell$+BICEP-DDM2. The dashed line is the best fit value, which coincides reasonably well with the marginalized values. }
   \label{fig:marginalized1}
\end{figure}
Although other models exist \cite{Liddle:1998jc,Dimopoulos2008}, a thoroughly investigated UV-complete model known as axion monodromy \cite{2005JCAP...01..005K,Silverstein2008,Mcallister2010,Flauger2010}, originally motivated by natural inflation \cite{PhysRevLett.65.3233}, can currently be considered as one of the few examples  \footnote{N-flation is an example that successfully avoids the Lyth bound, but this model is harder to implement in string theory \cite{Dimopoulos2008}} that addresses super Planckian  displacements. Derived within string theory, monodromy models ÒunwrapÓ angular directions in field space, and have monomial potentials over super-Planckian distance . The shift symmetry of the model is broken non-perturbatively to a discrete shift symmetry which results in an sinusodal feature in the potential \footnote{Strictly speaking, the symmetry is also broken by the slow-roll part of the potential, but this does not lead to oscillations.}. The resulting power spectrum can be approximated by a power law modified with logarithmically spaced osculations. In our analysis we will use the parametric form (scalars)\footnote{We use $\phi$ for the phase and $\phi_{\rm inf}$ for the inflaton where confusion might arise. }
\be
\Delta^2_{\mathcal{R}}(k)  &=& A_s \left(\frac{k}{k_*} \right)^{n_s-1}\left(1+\delta n_s \cos [\omega \log k/k_* +\phi] \right), \nonumber\\
 \label{eq:scalarspectra}
 \ee
 and for the tensors (see the appendix for a derivation)
\be
\Delta^2_{\mathcal{D}}(k)  &=& A_s \frac{r_*}{4} \left(\frac{k}{k_*} \right)^{-r_*/8}\left(1- \delta n_s \frac{r_*}{8\omega} \sin [\omega \log k/k_* +\phi]\right).\nonumber\\
 \label{eq:tensorspectra}
\ee
Specifically, for axion monodromy inflation $\omega = \sqrt{2\epsilon_*}/f = \sqrt{r_*}/(f\sqrt{8})$, with $f$ the axion decay constant. The oscillatory term in Eq.~\eqref{eq:tensorspectra} is generally subleading, since it suppressed by a factors of $r_*$, and one power of the frequency. For $f < M_p$ (note that $[f] = M_p^{\rm red}$) and $r\sim 10^{-1}$, we have $\omega \gtrsim 10^{-2}$. Hence, at best, such a level of primordial gravitional waves suggest oscillations in the tensor modes are suppressed with respect to those in the scalar spectrum by about an order of magnitude.  In fact, the suppression is always $\sqrt{r_*}f$, and hence one expects this term to be subdominant unless $f*M_p^{\rm red} > M_p$. 

\begin{figure*}[htbp] 
   \centering
   \includegraphics[width=7in]{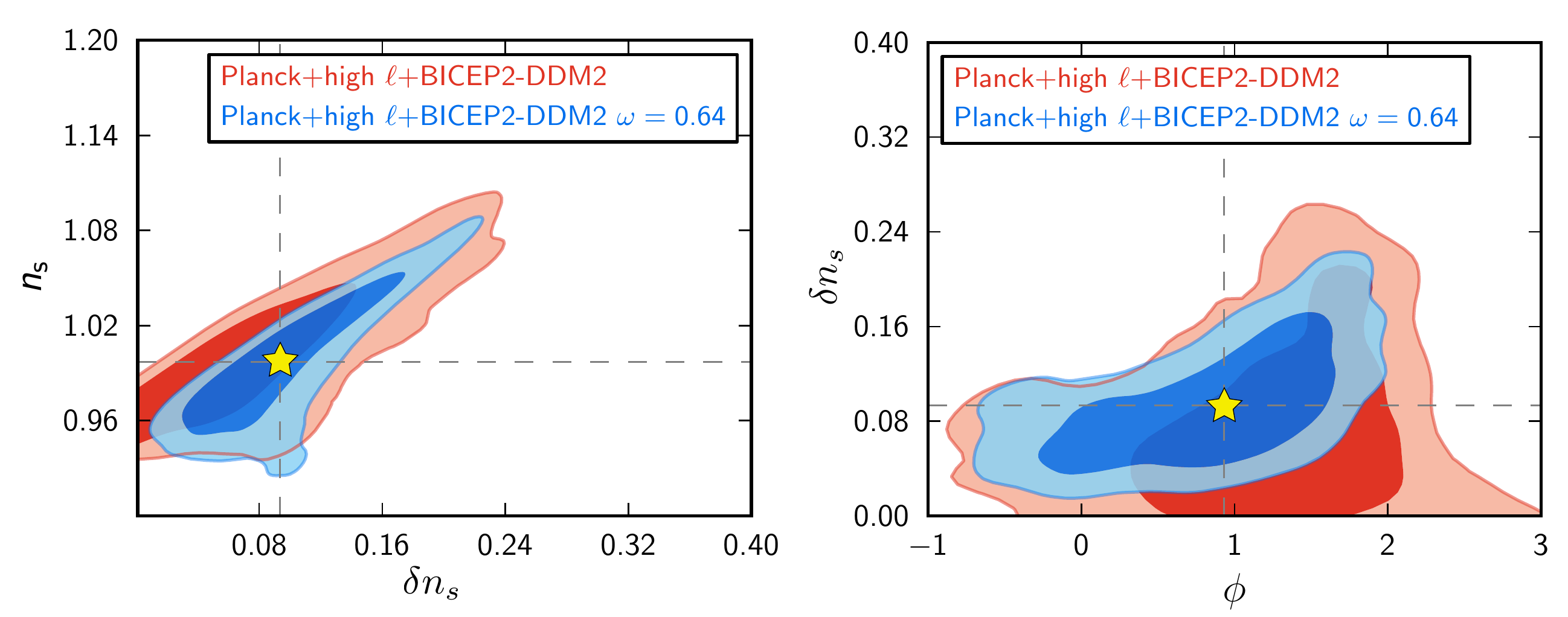} 
   \caption{2 Dimensional posterior probability distribution for $n_s$ vs $\delta n_s$ (left) and $\phi$ vs $\delta n_s$ (right). The stars represent the best fit points. The contours in red shades are derived from the chains with a prior on $\omega$ as specified in the text, while the blue shaded contours are derived from a chain with a set value for the frequency, close to the best fit value found using the previous chains. In the left it is clear that he tilt and amplitude are highly correlated; effectively the long wavelength oscillation acts a a tilt on low multipoles. }
   \label{fig:2d1}
\end{figure*}

Note that the above paramaterization also puts a constraint on the priors;  the spectrum could become negative if the amplitude were to become too large. The results presented here are derived in the assumption that the oscillations are small. We put a prior on the amplitude $0\leq \delta n_s \leq 0.3$, $0\leq r \leq 0.4$ and $0.01 \leq \omega \leq 2$, where the latter upper limit is somewhat arbitrary as axion monodromy does allow for higher frequency oscillations. Since we are only interested in at most a single oscillation to account for power suppression at large scales, we decided to cut off the frequency at this value, which roughly corresponds to 1 oscillation in the observed CMB. It can also be shown that oscillations with a frequency lower than this will be strongly correlated with other primordial parameters, while the correlation mostly disappears for higher frequencies, which favors a dedicated search varying all cosmological parameters at these low frequencies. 

This parametric form (without the tensor part) has been investigated in multiple publications \cite{Flauger2010,Meerburg2012,2013Flaugera,Meerburg2014b,Meerburg2014a,Meerburg2014c}, while \cite{Flauger2013} computed the exact numerical power spectrum directly from the potential. For further background on the details of axion monodromy inflation we would like to refer the reader to these excellent reviews \cite{PajerPeloso2013} and more generally \cite{DanLiam2014}. Other models that introduce logarithmic oscillations in the primordial power spectrum include e.g. models with a modified initial state \cite{Greene2005}, unwinding inflation \cite{Damico2013} and cascade inflation \cite{Ashoorioon2009}. Recent attempts to look for features in the CMB data can be found in e.g. Refs. \cite{Meerburg2012,2014PhRvD..89j3006A,2013JCAP...12..035H}.

\section{Results}\label{Results}

\begin{table*}
\centering
\begin{tabular}{| l | c | c | c | c | c | c | c | c | c | c | c | r |}
\hline\hline 
Data set  \textbackslash  Parameter & $\Omega_b h^2$ & $\Omega_c h^2$& $\tau$ & $H_0$ & $n_s $ &$\log 10^{10} A_s$ & $\delta n_s$ & $\phi$ & $ \omega$ & $r_{0.05}$& $\Delta \chi^2$ \tabularnewline
\hline 
Planck+high $\ell$ + BICEP2 & $0.022$& 0.119 & 0.096 & 67.54 & 1.004 & 3.082 & 0.0743 & 1.28 & 0.802 & 0.21 & $\sim 11$ \tabularnewline
Planck+high $\ell$ + BICEP2 - DDM2 & $0.022$& 0.119 & 0.098 & 67.74 & 0.997 & 3.076& 0.09 & 0.93 & 0.64 & 0.13 & $\sim 7$ \tabularnewline
\hline
\end{tabular}
\caption{Best-fit parameters for various data combinations. Note the tilt is $n_s\simeq1$ in order to account for effects on small scales due to the long wavelength oscillation. Subtracting potential foregrounds does not significantly change the parameter values, nor the significance. }
\label{tab:bestfits}
\centering
\end{table*}

In light of the observed level of B-modes, as well as the presence of a slight tension at large scales between the data and the theory, we performed an analysis of the BICEP2 data \cite{BICEP2014} in combination with the Planck \footnote{We use the low $\ell$ WMAP polarization data contained in Planck likelihood code.} public likelihood  \cite{planckLikelihood2013} and a modified version of the public cosmology package Cosmomc \cite{Lewis2002}. We also included small scale CMB data from the ACT \cite{ACT2014} and SPT \cite{SPT2011,SPT2012} experiments. Data from these experiments act as an extension of the leverage arm, suppressing large deviations from scale invariance (which is not preferred by these experiments). As we will see however, the best fit model does not affect the small scale power spectrum on CMB scales, but does affect the matter power spectrum, both on large scales and on small scales. 

Regarding the BICEP measurement, we will include a no-foreground, as well as foreground corrected analysis, where the latter is taken from the BICEP paper (the data-drives models, auto DDM1, and cross DDM2 spectra \cite{BICEP2014}). We use the standard Metropolis-Hasting sampling, and vary all parameters, including Planck foregrounds. Because of the high irregularity of the likelihood in the frequency direction, such a chain will never converge according to the $R-1$ convergence test. However, we terminate the chain by considering the convergence of each troubled parameter has reached $R-1 <0.1$. It must be noted, that the derived distributions are not a very accurate representation of the true distribution, as parts of the likelihood will be oversampled. The significance bounds we quote are derived from an analysis where we fixed the frequency to its best-fit value, derived from a sampling that included the frequency as a free parameters as described above. We also investigated using the nested sampler Multinest \cite{Feroz2009,Feroz2013}, however we found that including high $\ell$ data, oscillations and BICEP2 data resulted in very slow convergence. For our low $\ell$ analysis with fixed foregrounds we recover the same results as with the Metropolis-Hastings sampling, providing confidence in our obtained results using the MH sampler. Simply looking for the best-fit point also proofed troublesome, as existing methods are quite sensitive to the initial search volume.

Figure \ref{fig:marginalized1} shows an example of a 1D marginalized distribution derived from $Planck$ + high $\ell$+ $BICEP2$, where the latter has been corrected for foregrounds using the DDM2 dust model.  The dashed lines represent the best-fit values of the 4 parameters. We obtained the distribution for $\omega$ from varying all parameters, and the other 3 have been obtained by fixing the frequency to the best-fit value. We find $\delta n_s = 0.095 ^{+0.03}_{-0.05}$, $n_s = 1.0^{+0.03}_{-0.04}$, $\log 10^{10}A_s =3.06 \pm 0.04$ and $\phi = 0.85 ^{+0.9}_{-1.6}$. 

In Fig.~\ref{fig:2d1} we show the marginalized 2 dimensional contours between $n_s$ and the amplitude (left) and $\phi$ and the amplitude (right). The two contours in each plot are derived from the combination of Planck, SPT, ACT and foreground corrected BICEP data. The  figure on the left shows that the tilt and the amplitude are correlated, which is unsurprising since the long wavelength oscillation + associated amplitude can replace the scale dependence coming from the tilt. In Table \ref{tab:bestfits} we show the best-fit parameters derived from foreground cleaned and uncleaned analysis. Interestingly we find a best-fit value of $n_s\sim1$, i.e. close to a scale invariant form of the non-oscillating part of the potential. We will discuss this in the next section. 

\begin{table}
\centering
\begin{tabular}{| l | r |}
\hline\hline 
component & $\Delta \chi^2$ \tabularnewline
\hline 
Lowlike & $\sim -0.3 $  \tabularnewline
Lensing & $\sim 0.6 $ \tabularnewline
BICEP2 & $\sim -3.4 $ \tabularnewline
Commander & $\sim -8 $ \tabularnewline
CAMspec & $\sim 1.1 $ \tabularnewline
ACTSPT & $\sim -1.2 $ \tabularnewline
\hline
total & $\sim -11.2$ \tabularnewline
\hline
\end{tabular}
\caption{$\Delta \chi^2$ breakdown per component for the analysis of the axion model versus concordance model without foreground corrections. The improvement is most significant in commander and in BICEP2 as expected. Note also that there is small improvement in the high $\ell$ fit.  Including foreground subtraction, the improvement in commander and BICEP are lowered. }
\label{tab:deltachibreakdown}
\centering
\end{table}


\section{Discussion}\label{discussion}
\begin{figure}[htbp] 
   \centering
  \includegraphics[width=3.2in]{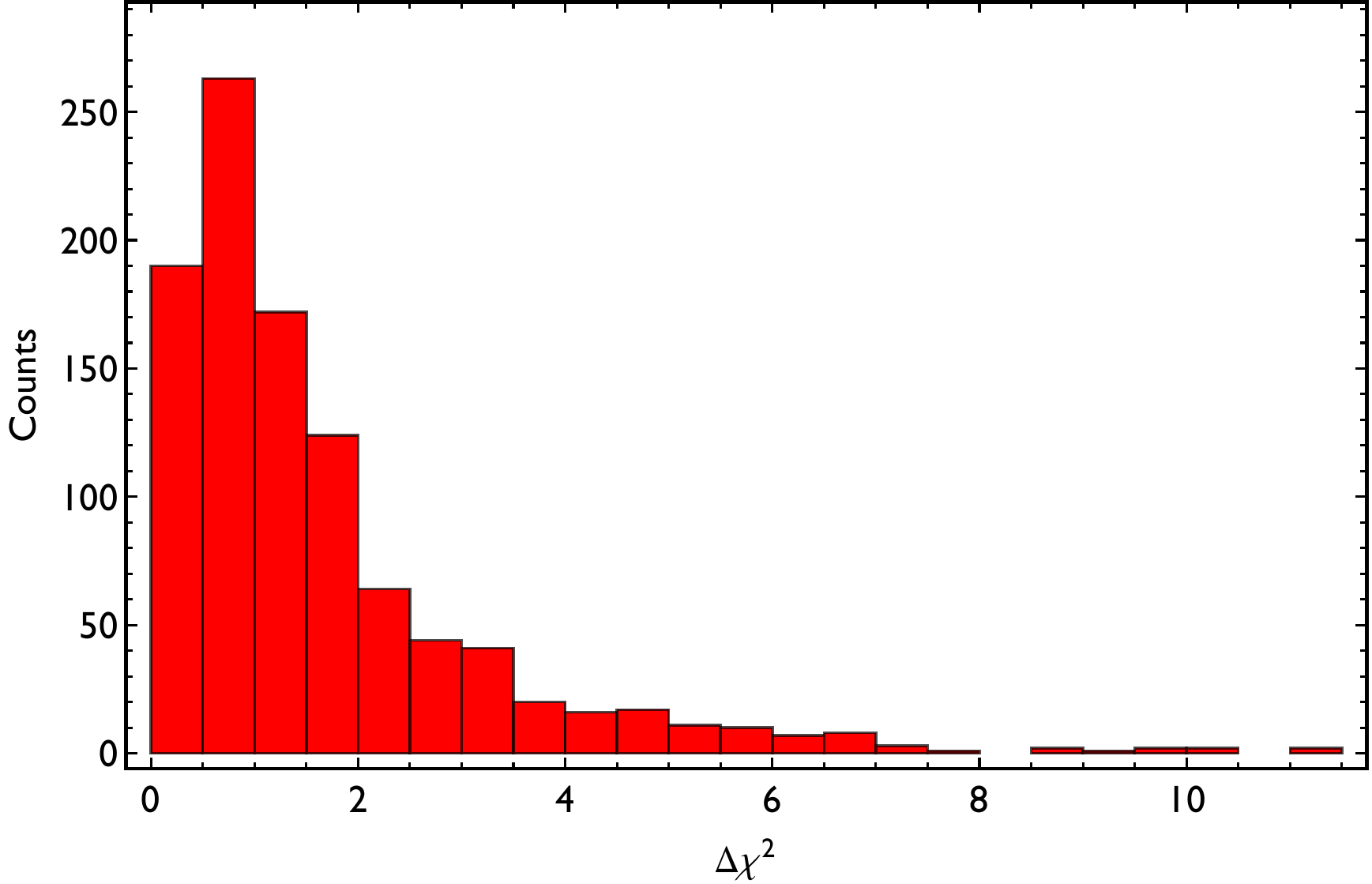} 
   \caption{Distribution of $\chi^2$ improvements simulating 1000 Universes and fitting for a feature on $\ell <1000$. This simple analysis suggest that cosmic variance can account for $\Delta \chi^2 \sim 2$. }
   \label{fig:distribution}
\end{figure}
\begin{figure*}[t] 
   \centering
   \includegraphics[width=3.25in]{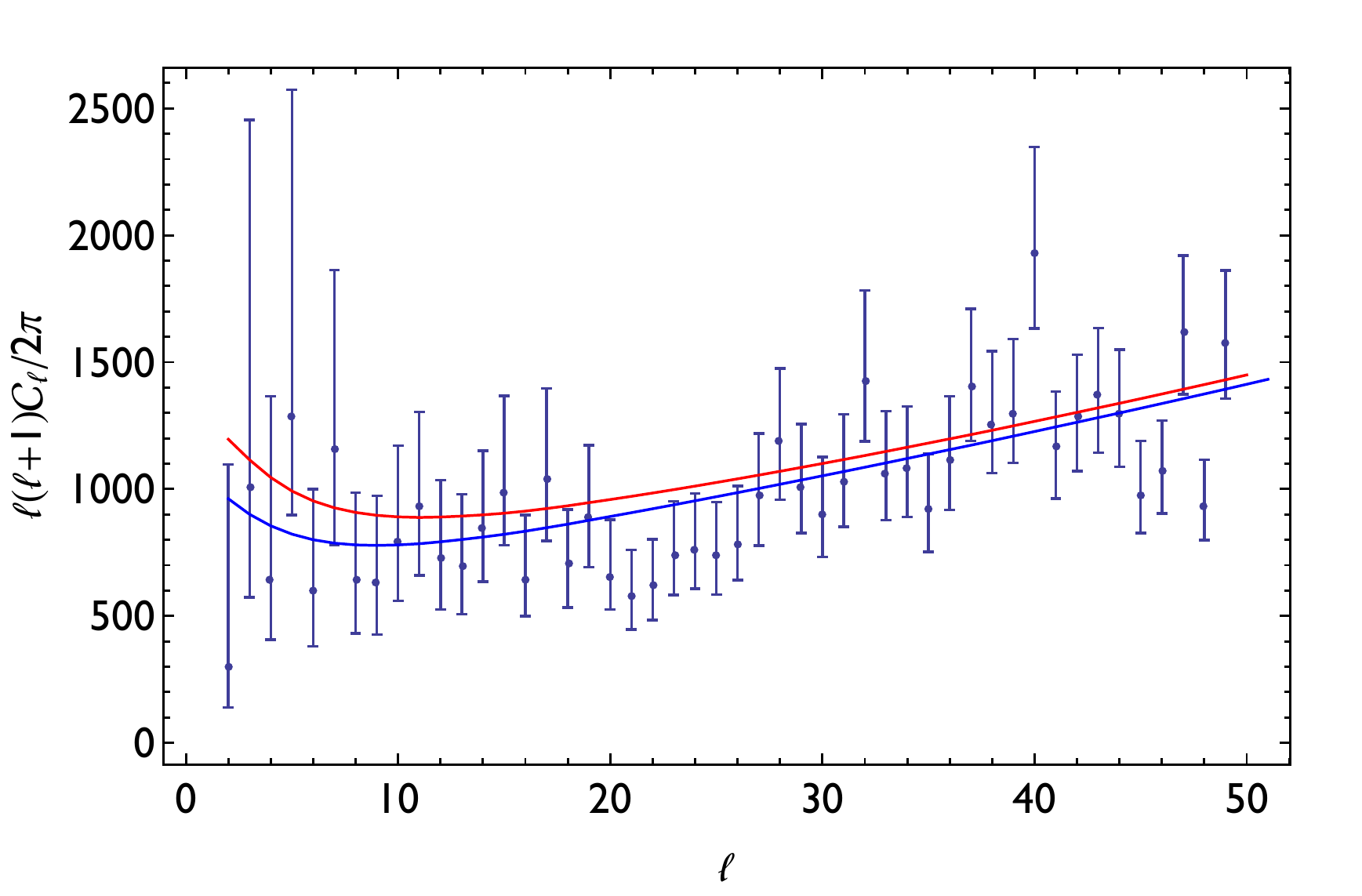} 
   \includegraphics[width=3.5in]{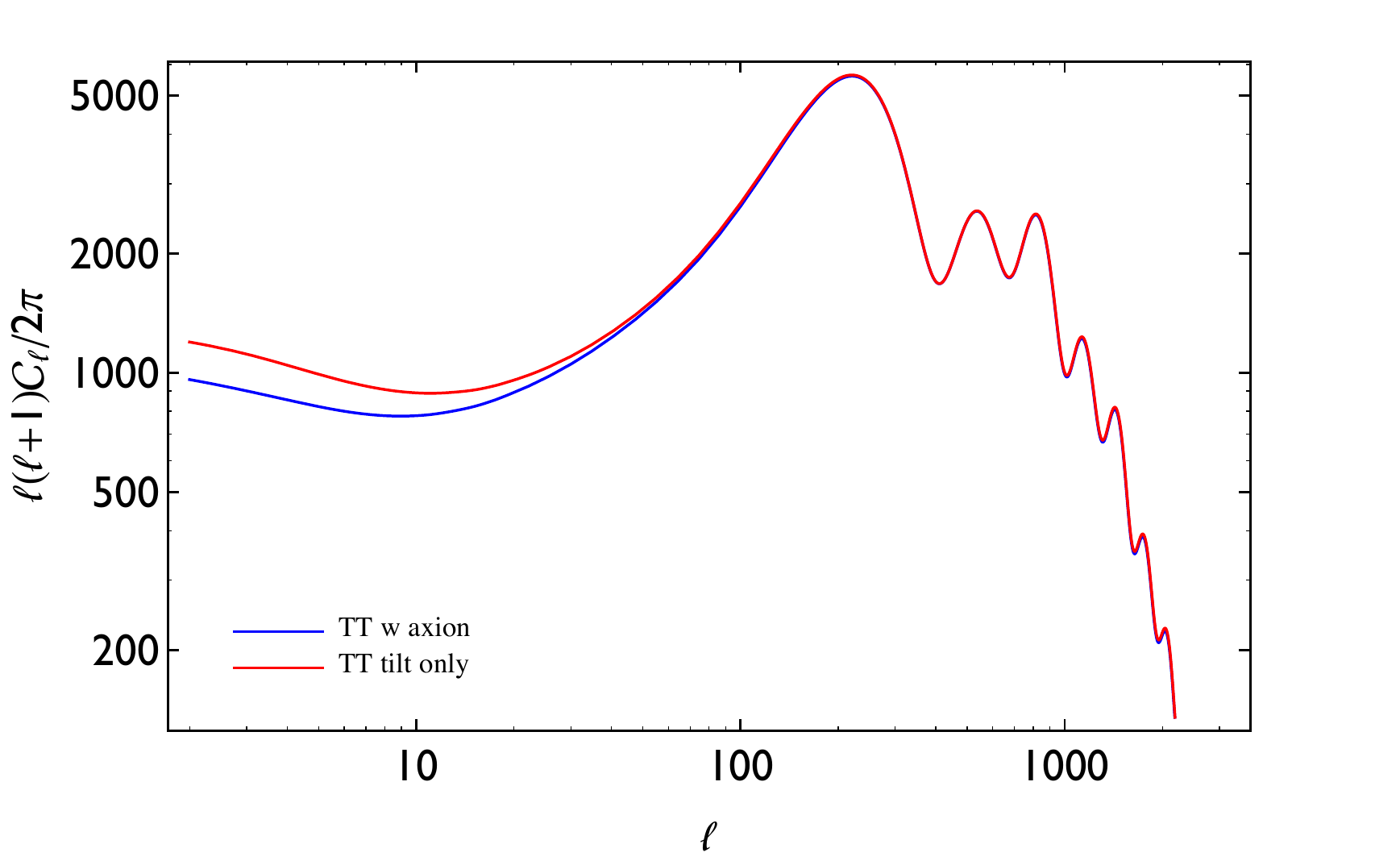} 
   \caption{The best fit TT spectrum  for a concordance cosmology (red) and the axion monodromy fitted model (blue). These plots are generated using the parameters derived from Planck+high $\ell$+lensing+WP+BICEP data.  }
   \label{fig:ClsTT}
\end{figure*}
In previous work \cite{2013Flaugera,Meerburg2014b} it was investigated to what extent the noise + cosmic variance can mimic a feature \footnote{In Ref. \cite{Meerburg2014c} we recently implemented our method into Multinest \cite{Feroz2009,Feroz2013}. }. It was shown that for a broad prior on the frequency, improvements are generally expected around $\Delta \chi^2=10$ making any improvement of this order in the data suspicious. The relatively small improvement here suggests we should worry about a fitting to the noise. However, as can be imagined, it becomes much harder to fit to the noise with a long wavelength feature, because points are correlated over a large range of scales (something that in principle the noise should not be). In Fig.~\ref{fig:distribution} we show the resulting distribution from a simple analysis where we try to fit an oscillation to the TT spectrum with cosmic variance an Planck-like noise, varying only the amplitude and the phase of a fixed feature. We only consider up to $\ell = 1000$, since we are interested in the improvement at low multipole. This simple analysis suggests that the noise alone can not account for a fit, given that the typical improvement from cosmic variance + noise os of the order $\Delta \chi^2 \sim2$. Unlike for high frequencies therefore, a  long wavelength modulation follows a typical $\chi^2$ distribution for $\sim 2$ parameters, which is not unexpected given the nature of the modulation (an effective rescaling of the amplitude). 

In Table~\ref{tab:deltachibreakdown} we break down the contribution to the improved $\Delta \chi^2$. The nature of the feature would suggest the improvement is dominated by large scales, hence by the commander component of the likelihood, which captures $\ell \leq 50$. It is reassuring to see that for the the contribution is indeed dominated by the commander likelihood as well as the BICEP2 likelihood (which causes the tension). At the same time, the likelihood at small scales is effectively unchanged.

\begin{figure*}[htbp] 
   \centering
   \includegraphics[width=3.5in]{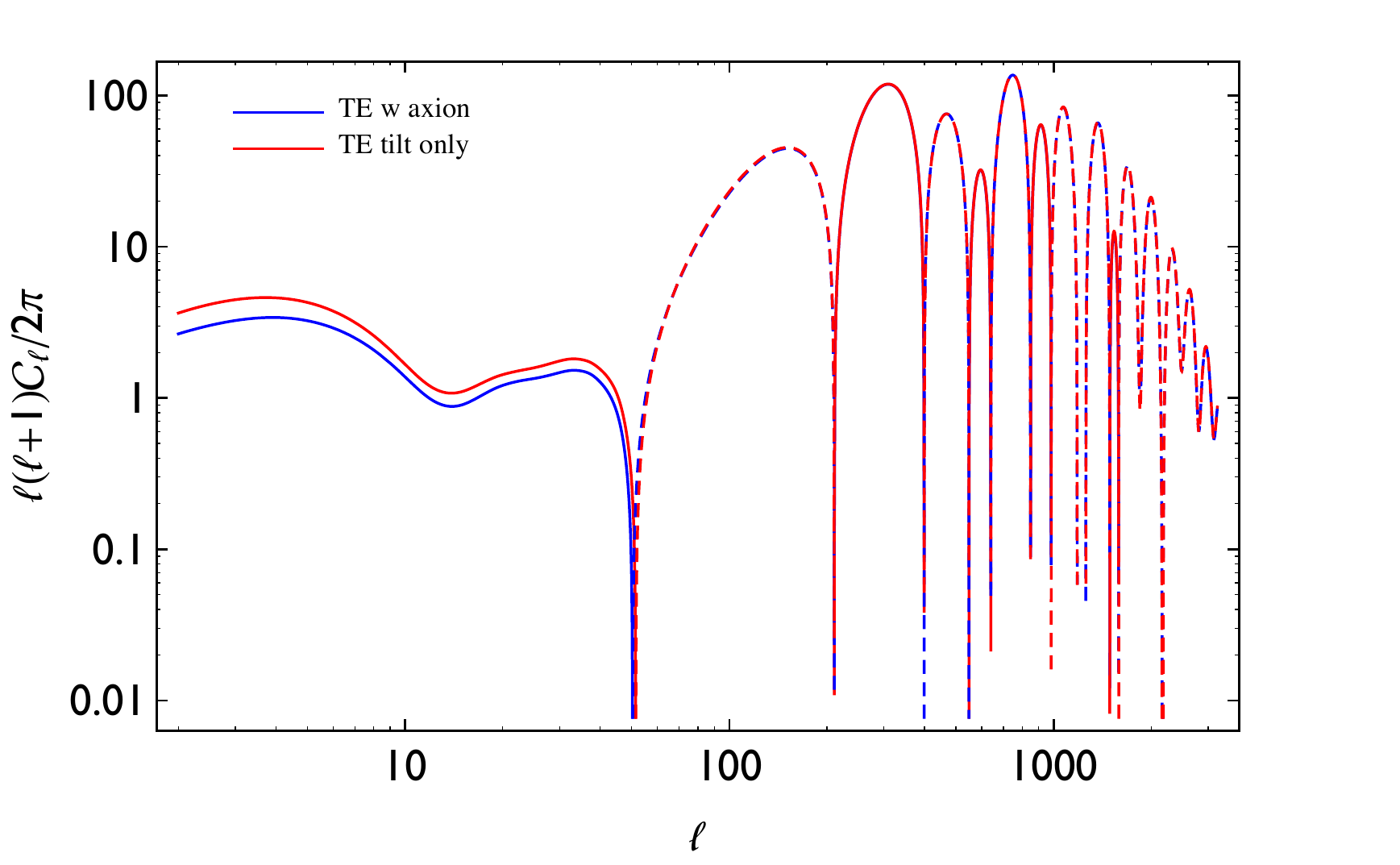} 
   \includegraphics[width=3.5in]{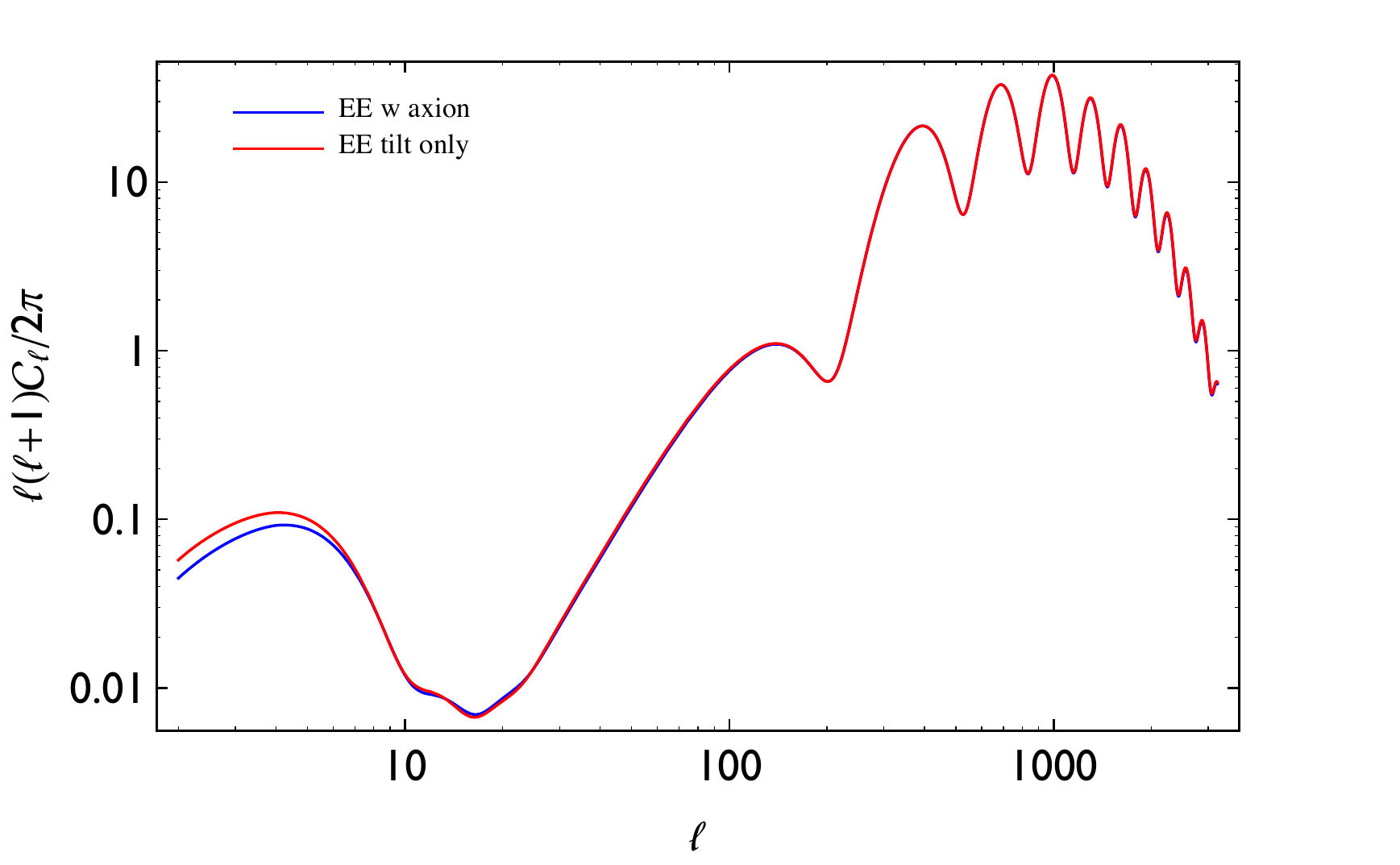} 
   \caption{The best fit TE (left) and TT(right) spectrum  for a concordance cosmology (red) and the axion monodromy fitted model (blue). These plots are generated using the parameters derived from Planck+high $\ell$+lensing+WP+BICEP data. For TE, the difference for the range $\ell =2-40$ is of order 20$\%$ for EE on the same range, the difference is about 10$\%$.  }
   \label{fig:ClsEETE}
\end{figure*}

In Fig.~\ref{fig:ClsTT} we show the best fit TT spectrum from this analysis and we compare this with the  best fit concordance cosmology. Quite interestingly, the resulting plot has a strong overlap at all $100\leq \ell \leq 3500$. The larger value of the tilt compensates for any deviations at large $\ell$ from the oscillation. In Fig.~\ref{fig:ClsEETE} we  show the anticipated difference with the EE and TE spectra. The difference between the concordance model for TE is relatively largest, and a Planck polarization measurements might be able to give further evidence for a large feature. These are relatively large scales, so they are cosmic variance limited. For TE over the range $\ell = 2-40$ the difference is about 20$\%$. We can estimate a detection by computing the signal to noise of the difference, i.e. for a full sky experiment
\be
\left(S/N\right)^2 \simeq \sum_{\ell = 2}^{40} (2\ell +1) \frac{(C^{\rm tilt}_{\ell}-C_{\ell}^{\rm axion})^2}{(C_{\ell}+N_{\ell})^2}
\ee
Given the difference of around $20\%$, and in the assumption we are limited by cosmic variance, we obtain $S/N\sim 8$, i.e. such a difference should be detectible. If one includes noise (which will be there for Planck) and the fact that for cosmology the sky fraction is about 30$\%$, we expect that Planck will be near the detection limit. What should be emphasized is that this simple estimate shown that a combined analysis of TT, TE and EE would help discriminate between the axion model and concordance cosmology. 

\begin{figure}[htbp] 
   \centering
  \includegraphics[width=3.5in]{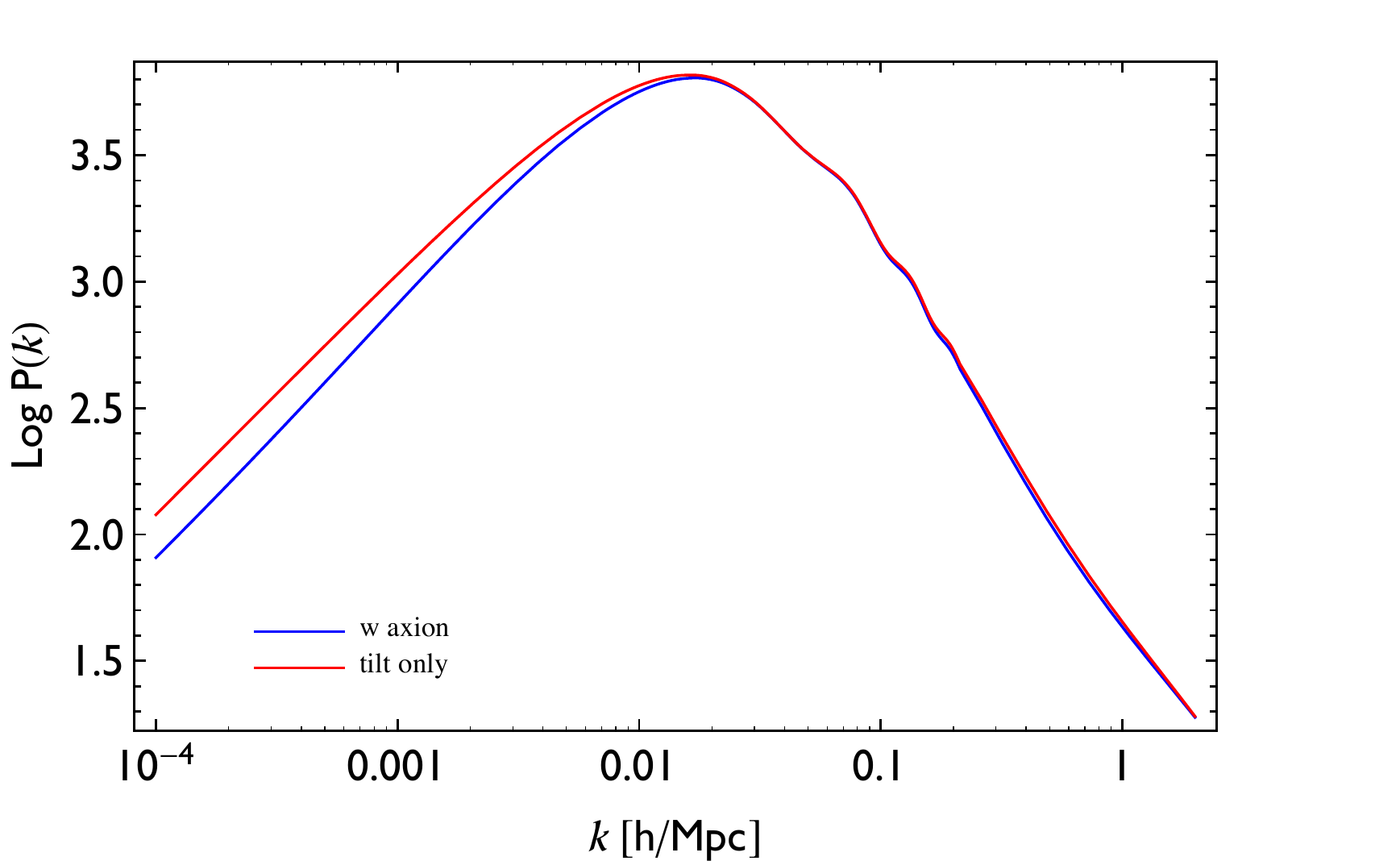} 
   \caption{The best fit matter spectrum for a concordance cosmology (red) and the axion monodromy fitted model (blue) at redshift $z=1.5$. These plots are generated using the parameters derived from Planck+high $\ell$+lensing+WP+BICEP data.  }
   \label{fig:matter power}
\end{figure}
\begin{figure*}[htbp] 
   \centering
  \includegraphics[width=3.5in]{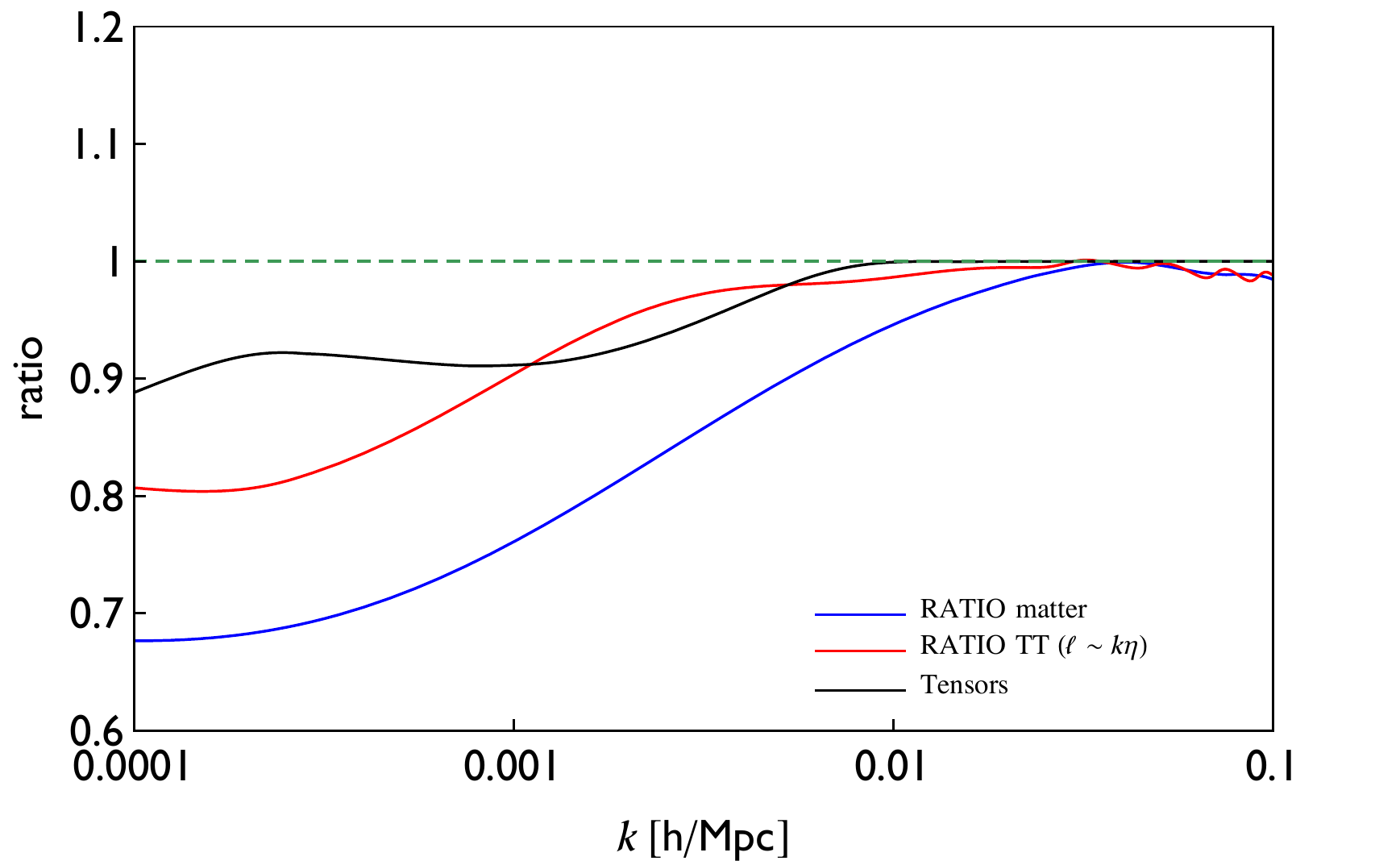} 
  \includegraphics[width=3.4in]{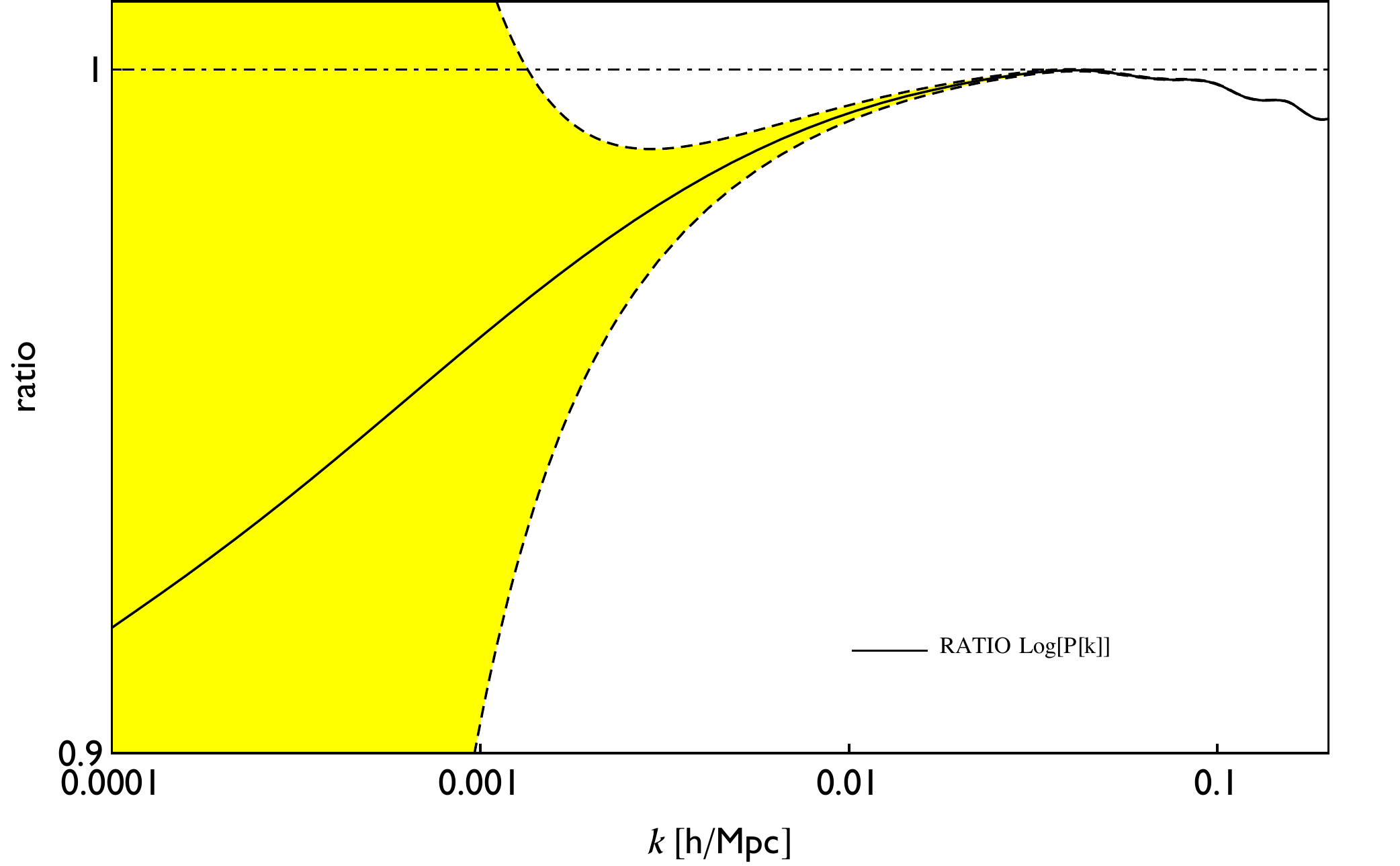} 
   \caption{(Left) The ratio of the measured power spectra with $r$ and $r+$ axion monodromy. We also show the contribution from the tensor modes only (which actually gives a positive contribution) with the aim of showing that the tensor are not contributing to the matter power spectrum and deviations from scale invariance of the form presented in this paper should be best constraint with a large scale measure of the matter power at redshift $z=1.5$. (Right) The projected detection of a feature in the large scale power spectrum with a spectroscopic galaxy survey with a volume $21$ Gpc$^3$ such as WFIRST \cite{WFIRST2013}. }
   \label{fig:ratio}
\end{figure*}

In Fig.~\ref{fig:matter power} we also show the matter power spectrum. At linear order the matter power spectrum is not affected by tensor modes, so the difference with the concordance model is much bigger as shown in the left panel of Fig.~\ref{fig:ratio}. In theory therefore, a measurement of the matter power spectrum at large scales would be the ideal way to constrain features on large scales. At very small scales, the model should start to deviate again, and a measure of small scales could further constrain a feature model. Interestingly, note that a feature also lowers the derived value of $\sigma_8$, putting it closer to the value estimated from $SZ$ measurements. We show the projected error of WFIRST on the right panel in Fig.~\ref{fig:ratio}. From this projection WFIRST \cite{WFIRST2013} should be able to distinguish a feature for modes $k\geq 10^{-2}$ Mpc$^{-1}$. 


\section{Conclusion} \label{Conclusion}

In this short paper we considered the novel possibility that a feature at large scales exists, alleviating marginal existing tension between the data and the $\Lambda$CDM concordance cosmology. Such a feature can be produced by various models, but in light of the super Planckian displacement of the field, the axion monodromy model appears to be a very interesting candidate. Although the predicted signal of such a model should produce a feature that persists even to the smallest scales, the logarithmic nature of the feature does not violate any measurements at small scales for very long wavelengths. We find that both $Planck+BICEP$ as well as the inclusion of small scale measurements of the CMB from $ACT$ and $SPT$, prefer such a feature and give an improvement of $\Delta \chi^2\sim 11$. This result is obtained from taking the BICEP data at face value and not correcting for foregrounds. We are however seriously concerned by the foregrounds and if we include the most optimistic foregrounds model, the improvement is reduced, but parameters are hardly affected. 

We have also shown that such a feature cannot be the result of cosmic variance, hence if the $BICEP$ findings are real, a feature of this type could persist. We discussed the potential of $Planck$ polarization measurements as a possible way to confirm of debunk such a feature in the near future. We also showed that because the matter power spectrum is not affected by tensors, the deviation at large scales from the concordance model would be bigger and WFIRST should be able to test the presence of such a feature with high significance. At the same time, a suppression of the matter power spectrum leads to a smaller value of the derived $\sigma_8$.  

The fitted feature introduces 3 additional parameters, we find that the best-fit tilt is close to 1, and the feature can take care of all the scale dependence within the observed window. Hence, one could argue, that at least within the observed window, one needs only 2 additional parameters. Obviously, this would require a model that naturally explains $n_s = 1$ (i.e. no power law scale dependence). We acknowledge that such a model would most likely be fine tuned and we should really count all 3 additional degrees of freedom if we want to address the significance of the improvement.  Quite surprisingly, the large tilt in combination with the feature, agrees excellently with the small scale measurements from $ACT$ and $SPT$, resulting to a spectrum that differs $< 2\%$  for $100 \leq \ell \leq 3200$ and does not worsen the fit at (the observed) small scales, a handicap of most models that try to alter the spectrum on large scales. 

The intriguing possibility exists that the data prefers an axion monodromy inflation scenario, with $f/M_p\simeq 0.01$ \footnote{A natural extension of the axion model includes a coupling to gauge field \cite{2011JCAP...04..009B}. One could combine the measurements here to put a constraint on the coupling \cite{2013JCAP...02..017M,2013PhRvD..87j3506L}, however it would require one to include the correction to the power spectrum from this coupling in the analysis which has not been done here. }. The model has the advantage of being UV complete and there is no need for fine-tuning. That being said, in principle the theoretical prior on the axion decay constant is weak and for completeness one should include all possible frequencies. Here we decided to focus on the low frequency end, since the new information in the form of a non-zero tensor-to scalar ratio will most likely not change the conclusion drawn in e.g. \cite{2013Flaugera,Meerburg2014b}; i.e. there is no compelling evidence for high oscillatory features in the CMB data. A problem with the current constraint is that the best-fit tensor to scalar ratio is rather large, and possibly too large for axion monodromy (although a new proposal results in larger values \cite{2014arXiv1405.3652M}). At the same time a tilt that close to $1$ is hard to achieve (although, naturally, for axion monodromy the tilt is closer to 1 than what is measured without a modulation). However, it is very plausible that future data, including a better understanding of the foregrounds will lower the value of $r$ and this would probably also lead to a tilt that is more consistent with axion monodromy potentials. This is already apparent from the simple analysis we have done here, where we subtracted data driven foreground models from the BB spectrum. The foregrounds subtraction indeed lowers $r$ and $n_s$ bringing them closer to theoretically predicted values. At the same time, the significance is lowered because the tension between the data and the concordance model is weakened. Ultimately, a full understanding of the foregrounds is necessary in order to address the  evidence of axion monodromy inflation, which could be a serious concern \cite{2014arXiv1405.7351F}. We hope to report on this in future work.

\section*{Acknowledgments}
The author would like to thank Raphael Flauger, David Spergel and Enrico Pajer for their contributions to this work. The author is part funded by the John Templeton Foundation grant number 37426.

\appendix 

\section{Scalar and tensor power spectra for large modulations.}
The axion monodromy potential is given by 
\be
V(\phi) &=& V_0(\phi)+\Lambda^4 \cos \left(\frac{\phi}{f} \right),
\ee
where $\phi$ is a canonically normalized scalar field and $V_0(\phi)$ is the slow-roll potential in the absence of modulations. A model with a linear potential has been derived from string theory, but for the remainder of this derivation will not be specifically considered. $\Lambda$ and $f$ have the dimensions of mass. One can define a monoticity parameter
\be
b_* \equiv \frac{\Lambda^4}{V'(\phi_*)f}<1,
\ee
which sets the requirement that the potential is monotonic for the scales observed the the CMB. The monoticity parameter is computed  at $\phi = \phi_*$ when the mode $k = k_*$ exists the horizon. We will be using the following Hubble slow-roll parameters
\be
\epsilon\equiv - \frac{\dot{H}}{H^2} \; \; \; \delta \equiv \frac{\ddot{H}}{2\dot{H}H}.
\ee
One can derive a the time dependence of the scalar field from the potential
\be
\phi(t) &=&\phi_0(t) -\frac{3b_*f^2}{\sqrt{2\epsilon_*}}\sin \left(\frac{\phi_0(t)}{f}\right),
\ee
where $\epsilon_*$ is the slow-roll parameter in the absence of modulations evaluated at $\phi_0 = \phi_*$. We expend the slow-roll parameters in power of the monoticity parameter (which has to be small)
\be
\epsilon = \epsilon_0 + \epsilon_1 +\mathcal{O}(b_*^2),\\
\delta = \delta_0 + \delta_1 +\mathcal{O}(b_*^2).
\ee
In the slow-roll approximation for the unmodulated field $\phi_0$ and 
\be
\epsilon_0 &=& \epsilon_*\\
\epsilon_1 &=& -3b_* \frac{f\sqrt{2\epsilon_*}}{1+(3f/\sqrt{2\epsilon_*})^2} \left[\cos\left( \frac{\phi_0}{f}\right) + \right. \nonumber \\ 
&& \left. (3f/ \sqrt{3 \epsilon_*})\sin \left( \frac{\phi_0}{f}\right) \right],\\
\delta_1 &=&  -3b_* \frac{1}{1+(3f/\sqrt{2\epsilon_*})^2} \left[\sin\left( \frac{\phi_0}{f}\right) - \right. \nonumber \\ 
&& \left. (3f/ \sqrt{3 \epsilon_*})\cos \left( \frac{\phi_0}{f}\right) \right].
\ee

One can compute the scalar perturbations and the tensor perturbations as follows. We pick the following gauge 
\be
\delta g_{ij}({\bf x},t)&=& 2a(t)^2 \R({\bf x},t)\delta_{ij},
\ee
and 
\be
\R({\bf x},t) &=& \int \frac{d^3{\bf k}}{(2\pi)^3} \R({\bf k}, t) e^{i {\bf k} \cdot {\bf x}}.
\ee
The most general solution $\R$ in case of  rotational invariance and hermiticity can be written as 
\be
\R({\bf k}, t)&=& \R_{k} a({\bf k}) + \R_k^* a^{\dagger} (-{\bf k}),
\ee
where the creation and annihilation operators satisfy the usual commutation relations 
\be
\left[ a({\bf k}) ,a^{\dagger} ({\bf k}')\right] &=& (2 \pi)^3 \delta({\bf k}-{\bf k}').
\ee

The Mukhanov-Sasaki equation can be written as 
\be
\frac{d^2 \R_k}{dx^2}-\frac{2(1+2\epsilon +\delta)}{x} \frac{d \R_k}{d x} +\R_k &=&0,
\ee
with $x = -k\tau$ and $\tau$ the conformal time. In de Sitter, the infinite past lies at $\tau \rightarrow -\infty$ and the initial conditions are such that 
\be
\R_k(x) \rightarrow -\frac{H}{\sqrt{2k}a\dot{\phi}}e^{ix}.
\ee
In the future, $\tau \rightarrow 0$ and the mode functions $R_k$ approach a constant which will be donated by $\R^{(0)}_k$. The associated spectrum of perturbations is related to $\R_k$ as 
\be
\left| \R^{(0)}_k\right|^2 = (2\pi)^2 \frac{\Delta_{\R}^2(k)}{k^3}.
\label{eq:nomod}
\ee
For tensors we can derive a similar equation \cite{steven2008cosmology}
\be
\frac{d^2 \D_k}{dx^2}-2\frac{1+\epsilon}{x} \frac{d \D_k}{d x} +\D_k &=&0.
\ee
In \cite{Flauger2010} it was shown that the quantity of interest to first order in $b_*$ is given by
\be
\left| \R^{(0)}_k\right|^2  & = & \left| \R^{(0)}_{k,0}\right|^2 \left[1+2{\rm Re}g_{1}(0)\right],
\ee
for scalars and 
\be
\left| \D^{(0)}_k\right|^2  & = & \left| \D^{(0)}_{k,0}\right|^2 \left[1+2{\rm Re}g_2(0)\right],
\ee
for tensors. $ \left| \R^{(0)}_{k,0}\right|^2$ and $\left| \D^{(0)}_{k,0}\right|^2$ are the solutions without modulations as in Eq. \eqref{eq:nomod}, and the correction obey the Mukhanov-Sasaki equations
\be
\frac{d^2g_{1,2}}{dx^2} -\frac{2}{x}\frac{dg_{1,2}}{dx}+g_{1,2} = 2e^{ix} S_{1,2},
\ee
with $S_1 = 2\epsilon_1+ \delta_1$ for scalars and $S_2= \epsilon_1$ for tensors. We will work to leading order in slow-roll for tensors and scalars, dropping the contribution from $\epsilon$ for scalars.

Using trigonometric identities, one can show that 
\be
\epsilon_1 &=& \frac{-3b_* f \sqrt{2\epsilon_*}}{\sqrt{1+(3 f /{\sqrt{2\epsilon_*})^2}}} \sin \left(\frac{\phi_0(x)}{f} + \psi\right), \\
\delta_1 &= & \frac{3b_* }{\sqrt{1+(3 f /{\sqrt{2\epsilon_*})^2}}} \cos \left(\frac{\phi_0(x)}{f} + \psi\right).
\ee
with $\psi = \arcsin \left[(1+(3 f /\sqrt{2\epsilon_*})^2)^{-1/2}\right]$. Here we will neglect this phase, since it will be unimportant and in the search for modulations in the data, we only care about the relative phase between the tensor and scalar modulations. 

Using the relation $\phi_0(x) = (2\epsilon_*)^{-1/2} -(2\epsilon_*)^{1/2} \ln (k/k_*)= \phi_k + (2\epsilon_*)^{1/2} \ln x$, and $r_{1,2}(x) \equiv {\rm Re} (g_{1,2}(x))$ we write the Mukhanov-Sasaki equations for the scalars as 
\begin{widetext}
\be
\frac{d^2 r_1}{dx^2}-\frac{2}{x}\frac{dr_1}{dx} + r_1 = \frac{6b_*}{\sqrt{1+(3 f /{\sqrt{2\epsilon_*})^2}}} \cos(x) \cos\left(\phi_k/f + \sqrt{2\epsilon_*} \ln x/f\right), 
\ee 
and for tensors 
\be
\frac{d^2 r_2}{dx^2}-\frac{2}{x}\frac{dr_2}{dx} + r_2 = \frac{-6b_* f \sqrt{2\epsilon_*}}{\sqrt{1+(3 f /{\sqrt{2\epsilon_*})^2}}} \cos(x) \sin\left(\phi_k/f +  \sqrt{2\epsilon_*} \ln x/f\right).  
\ee
\end{widetext}

The solutions can be found using Green's functions. We are interested in the solution at late times when $x\rightarrow 0$. Following \cite{Flauger2010} and \cite{FlaugerPajer2011} we find up to a phase
\be
r_1(0) = \delta n_s \cos \left( \frac{\phi_k}{f}\right), 
\ee
and 
\be
r_2(0) = - f \sqrt{2\epsilon_*}\delta n_s \sin \left( \frac{\phi_k}{f}\right), 
\ee
with 
\be
\delta n_s &=& \frac{12 b_*}{\sqrt{1+(3 f /{\sqrt{2\epsilon_*})^2}}} \sqrt{\frac{\pi}{8} \coth \left( \frac{\pi \sqrt{2\epsilon_*}}{2f}\right) \frac{f}{\sqrt{2\epsilon_*}}}.\nonumber\\
\ee
Note that compared to the scalar power spectrum, the tensor modulation is suppressed in amplitude with a factor $\frac{\sqrt{2 \epsilon_*}}{f} > 1$. We can use the definition of the tensor to scalar ratio (note there are 2 helicity modes for the tensors)
\be
r_k \equiv 4 \left| \D^{(0)}_k\right|^2/\left| \R^{(0)}_k\right|^2,
\ee
and for the power law of the potential $r = 16 \epsilon$ or, for a fixed pivot scale $r_* = 16 \epsilon_*$. We can thus write down our power spectra as 
\be
\Delta_{\R}^2(k) = \Delta_{\R}^2(k_*) \left(\frac{k}{k_*}\right)^{n_s-1} \left[1+\delta n_s \cos \left(\frac{\phi_k}{f}\right) \right], \nonumber \\
\ee
and 
\be
\Delta_{\D}^2(k) = \frac{r_*}{4} \Delta_{\R}^2(k_*)\left(\frac{k}{k_*}\right)^{-r_*/8}  \left[1-\delta n_s \frac{(r_*^{1/2}f)}{(8)^{1/2}}\sin \left(\frac{\phi_k}{f}\right) \right]. \nonumber \\
\ee

We can use these expressions to build consistent templates
\be
\Delta^2_{\mathcal{R}}(k)  &=& A_s \left(\frac{k}{k_*} \right)^{n_s-1}\left(1+\delta n_s \cos [\omega \log k/k_* +\phi] \right),\nonumber\\
 \label{eq:powerspectra1}
 \Delta^2_{\mathcal{D}}(k)  &=& A_s \frac{r_*}{4} \left(\frac{k}{k_*} \right)^{-r_*/8}\left(1- \delta n_s \frac{r_*}{8\omega} \sin [\omega \log k/k_* +\phi]\right).\nonumber
\ee

Note that the term is subleading term in the scalars (i.e. $\propto \epsilon$) is of the same order as the tensors. However, in the derivation of the equation of motion, terms of the same order have been dropped and these terms are suppressed by almost 2 orders of magnitude for $\omega \sim 1$ (the frequency range of interest for a long wavelength modulation). 

\bibliography{osc_paper}

\begin{thebibliography}{47}
\expandafter\ifx\csname natexlab\endcsname\relax\def\natexlab#1{#1}\fi
\expandafter\ifx\csname bibnamefont\endcsname\relax
  \def\bibnamefont#1{#1}\fi
\expandafter\ifx\csname bibfnamefont\endcsname\relax
  \def\bibfnamefont#1{#1}\fi
\expandafter\ifx\csname citenamefont\endcsname\relax
  \def\citenamefont#1{#1}\fi
\expandafter\ifx\csname url\endcsname\relax
  \def\url#1{\texttt{#1}}\fi
\expandafter\ifx\csname urlprefix\endcsname\relax\def\urlprefix{URL }\fi
\providecommand{\bibinfo}[2]{#2}
\providecommand{\eprint}[2][]{\url{#2}}

\bibitem[{\citenamefont{{BICEP2 Collaboration}
  et~al.}(2014)\citenamefont{{BICEP2 Collaboration}, {Ade}, {Aikin}, {Barkats},
  {Benton}, {Bischoff}, {Bock}, {Brevik}, {Buder}, {Bullock}
  et~al.}}]{BICEP2014}
\bibinfo{author}{\bibnamefont{{BICEP2 Collaboration}}},
  \bibinfo{author}{\bibfnamefont{P.~A.~R.} \bibnamefont{{Ade}}},
  \bibinfo{author}{\bibfnamefont{R.~W.} \bibnamefont{{Aikin}}},
  \bibinfo{author}{\bibfnamefont{D.}~\bibnamefont{{Barkats}}},
  \bibinfo{author}{\bibfnamefont{S.~J.} \bibnamefont{{Benton}}},
  \bibinfo{author}{\bibfnamefont{C.~A.} \bibnamefont{{Bischoff}}},
  \bibinfo{author}{\bibfnamefont{J.~J.} \bibnamefont{{Bock}}},
  \bibinfo{author}{\bibfnamefont{J.~A.} \bibnamefont{{Brevik}}},
  \bibinfo{author}{\bibfnamefont{I.}~\bibnamefont{{Buder}}},
  \bibinfo{author}{\bibfnamefont{E.}~\bibnamefont{{Bullock}}},
  \bibnamefont{et~al.}, \bibinfo{journal}{ArXiv e-prints}
  (\bibinfo{year}{2014}), \eprint{1403.3985}.

\bibitem[{\citenamefont{{Planck Collaboration}
  et~al.}(2013{\natexlab{a}})\citenamefont{{Planck Collaboration}, {Ade},
  {Aghanim}, {Armitage-Caplan}, {Arnaud}, {Ashdown}, {Atrio-Barandela},
  {Aumont}, {Baccigalupi}, {Banday} et~al.}}]{planckcosmoparms2013}
\bibinfo{author}{\bibnamefont{{Planck Collaboration}}},
  \bibinfo{author}{\bibfnamefont{P.~A.~R.} \bibnamefont{{Ade}}},
  \bibinfo{author}{\bibfnamefont{N.}~\bibnamefont{{Aghanim}}},
  \bibinfo{author}{\bibfnamefont{C.}~\bibnamefont{{Armitage-Caplan}}},
  \bibinfo{author}{\bibfnamefont{M.}~\bibnamefont{{Arnaud}}},
  \bibinfo{author}{\bibfnamefont{M.}~\bibnamefont{{Ashdown}}},
  \bibinfo{author}{\bibfnamefont{F.}~\bibnamefont{{Atrio-Barandela}}},
  \bibinfo{author}{\bibfnamefont{J.}~\bibnamefont{{Aumont}}},
  \bibinfo{author}{\bibfnamefont{C.}~\bibnamefont{{Baccigalupi}}},
  \bibinfo{author}{\bibfnamefont{A.~J.} \bibnamefont{{Banday}}},
  \bibnamefont{et~al.}, \bibinfo{journal}{ArXiv e-prints}
  (\bibinfo{year}{2013}{\natexlab{a}}), \eprint{1303.5076}.

\bibitem[{\citenamefont{{Abazajian} et~al.}(2014)\citenamefont{{Abazajian},
  {Aslanyan}, {Easther}, and {Price}}}]{Abazajian2014}
\bibinfo{author}{\bibfnamefont{K.~N.} \bibnamefont{{Abazajian}}},
  \bibinfo{author}{\bibfnamefont{G.}~\bibnamefont{{Aslanyan}}},
  \bibinfo{author}{\bibfnamefont{R.}~\bibnamefont{{Easther}}},
  \bibnamefont{and} \bibinfo{author}{\bibfnamefont{L.~C.}
  \bibnamefont{{Price}}}, \bibinfo{journal}{ArXiv e-prints}
  (\bibinfo{year}{2014}), \eprint{1403.5922}.

\bibitem[{\citenamefont{{Hazra} et~al.}(2014)\citenamefont{{Hazra},
  {Shafieloo}, {Smoot}, and {Starobinsky}}}]{Hazra2014}
\bibinfo{author}{\bibfnamefont{D.~K.} \bibnamefont{{Hazra}}},
  \bibinfo{author}{\bibfnamefont{A.}~\bibnamefont{{Shafieloo}}},
  \bibinfo{author}{\bibfnamefont{G.~F.} \bibnamefont{{Smoot}}},
  \bibnamefont{and} \bibinfo{author}{\bibfnamefont{A.~A.}
  \bibnamefont{{Starobinsky}}}, \bibinfo{journal}{ArXiv e-prints}
  (\bibinfo{year}{2014}), \eprint{1404.0360}.

\bibitem[{\citenamefont{{Planck Collaboration}
  et~al.}(2013{\natexlab{b}})\citenamefont{{Planck Collaboration}, {Ade},
  {Aghanim}, {Armitage-Caplan}, {Arnaud}, {Ashdown}, {Atrio-Barandela},
  {Aumont}, {Baccigalupi}, {Banday} et~al.}}]{PlanckSZ2013}
\bibinfo{author}{\bibnamefont{{Planck Collaboration}}},
  \bibinfo{author}{\bibfnamefont{P.~A.~R.} \bibnamefont{{Ade}}},
  \bibinfo{author}{\bibfnamefont{N.}~\bibnamefont{{Aghanim}}},
  \bibinfo{author}{\bibfnamefont{C.}~\bibnamefont{{Armitage-Caplan}}},
  \bibinfo{author}{\bibfnamefont{M.}~\bibnamefont{{Arnaud}}},
  \bibinfo{author}{\bibfnamefont{M.}~\bibnamefont{{Ashdown}}},
  \bibinfo{author}{\bibfnamefont{F.}~\bibnamefont{{Atrio-Barandela}}},
  \bibinfo{author}{\bibfnamefont{J.}~\bibnamefont{{Aumont}}},
  \bibinfo{author}{\bibfnamefont{C.}~\bibnamefont{{Baccigalupi}}},
  \bibinfo{author}{\bibfnamefont{A.~J.} \bibnamefont{{Banday}}},
  \bibnamefont{et~al.}, \bibinfo{journal}{ArXiv e-prints}
  (\bibinfo{year}{2013}{\natexlab{b}}), \eprint{1303.5080}.

\bibitem[{\citenamefont{{Flauger} et~al.}(2014)\citenamefont{{Flauger}, {Hill},
  and {Spergel}}}]{2014arXiv1405.7351F}
\bibinfo{author}{\bibfnamefont{R.}~\bibnamefont{{Flauger}}},
  \bibinfo{author}{\bibfnamefont{J.~C.} \bibnamefont{{Hill}}},
  \bibnamefont{and} \bibinfo{author}{\bibfnamefont{D.~N.}
  \bibnamefont{{Spergel}}}, \bibinfo{journal}{ArXiv e-prints}
  (\bibinfo{year}{2014}), \eprint{1405.7351}.

\bibitem[{\citenamefont{{Lyth} and {Riotto}}(1999)}]{Lyth1999}
\bibinfo{author}{\bibfnamefont{D.~H.} \bibnamefont{{Lyth}}} \bibnamefont{and}
  \bibinfo{author}{\bibfnamefont{.~A.} \bibnamefont{{Riotto}}},
  \bibinfo{journal}{\physrep} \textbf{\bibinfo{volume}{314}},
  \bibinfo{pages}{1} (\bibinfo{year}{1999}), \eprint{hep-ph/9807278}.

\bibitem[{\citenamefont{{Easther} et~al.}(2006)\citenamefont{{Easther},
  {Kinney}, and {Powell}}}]{EastherKinney2006}
\bibinfo{author}{\bibfnamefont{R.}~\bibnamefont{{Easther}}},
  \bibinfo{author}{\bibfnamefont{W.~H.} \bibnamefont{{Kinney}}},
  \bibnamefont{and} \bibinfo{author}{\bibfnamefont{B.~A.}
  \bibnamefont{{Powell}}}, \bibinfo{journal}{\jcap}
  \textbf{\bibinfo{volume}{8}}, \bibinfo{eid}{004} (\bibinfo{year}{2006}),
  \eprint{astro-ph/0601276}.

\bibitem[{\citenamefont{{Baumann} and {Green}}(2012)}]{BaumannGreen2012}
\bibinfo{author}{\bibfnamefont{D.}~\bibnamefont{{Baumann}}} \bibnamefont{and}
  \bibinfo{author}{\bibfnamefont{D.}~\bibnamefont{{Green}}},
  \bibinfo{journal}{\jcap} \textbf{\bibinfo{volume}{5}}, \bibinfo{eid}{017}
  (\bibinfo{year}{2012}), \eprint{1111.3040}.

\bibitem[{\citenamefont{Liddle et~al.}(1998)\citenamefont{Liddle, Mazumdar, and
  Schunck}}]{Liddle:1998jc}
\bibinfo{author}{\bibfnamefont{A.~R.} \bibnamefont{Liddle}},
  \bibinfo{author}{\bibfnamefont{A.}~\bibnamefont{Mazumdar}}, \bibnamefont{and}
  \bibinfo{author}{\bibfnamefont{F.~E.} \bibnamefont{Schunck}},
  \bibinfo{journal}{Phys.Rev.} \textbf{\bibinfo{volume}{D58}},
  \bibinfo{pages}{061301} (\bibinfo{year}{1998}), \eprint{astro-ph/9804177}.

\bibitem[{\citenamefont{{Dimopoulos} et~al.}(2008)\citenamefont{{Dimopoulos},
  {Kachru}, {McGreevy}, and {Wacker}}}]{Dimopoulos2008}
\bibinfo{author}{\bibfnamefont{S.}~\bibnamefont{{Dimopoulos}}},
  \bibinfo{author}{\bibfnamefont{S.}~\bibnamefont{{Kachru}}},
  \bibinfo{author}{\bibfnamefont{J.}~\bibnamefont{{McGreevy}}},
  \bibnamefont{and} \bibinfo{author}{\bibfnamefont{J.~G.}
  \bibnamefont{{Wacker}}}, \bibinfo{journal}{\jcap}
  \textbf{\bibinfo{volume}{8}}, \bibinfo{eid}{003} (\bibinfo{year}{2008}),
  \eprint{hep-th/0507205}.

\bibitem[{\citenamefont{{Kim} et~al.}(2005)\citenamefont{{Kim}, {Nilles}, and
  {Peloso}}}]{2005JCAP...01..005K}
\bibinfo{author}{\bibfnamefont{J.~E.} \bibnamefont{{Kim}}},
  \bibinfo{author}{\bibfnamefont{H.~P.} \bibnamefont{{Nilles}}},
  \bibnamefont{and} \bibinfo{author}{\bibfnamefont{M.}~\bibnamefont{{Peloso}}},
  \bibinfo{journal}{\jcap} \textbf{\bibinfo{volume}{1}}, \bibinfo{eid}{005}
  (\bibinfo{year}{2005}), \eprint{hep-ph/0409138}.

\bibitem[{\citenamefont{{Silverstein} and {Westphal}}(2008)}]{Silverstein2008}
\bibinfo{author}{\bibfnamefont{E.}~\bibnamefont{{Silverstein}}}
  \bibnamefont{and}
  \bibinfo{author}{\bibfnamefont{A.}~\bibnamefont{{Westphal}}},
  \bibinfo{journal}{\prd} \textbf{\bibinfo{volume}{78}}, \bibinfo{eid}{106003}
  (\bibinfo{year}{2008}), \eprint{0803.3085}.

\bibitem[{\citenamefont{{McAllister} et~al.}(2010)\citenamefont{{McAllister},
  {Silverstein}, and {Westphal}}}]{Mcallister2010}
\bibinfo{author}{\bibfnamefont{L.}~\bibnamefont{{McAllister}}},
  \bibinfo{author}{\bibfnamefont{E.}~\bibnamefont{{Silverstein}}},
  \bibnamefont{and}
  \bibinfo{author}{\bibfnamefont{A.}~\bibnamefont{{Westphal}}},
  \bibinfo{journal}{\prd} \textbf{\bibinfo{volume}{82}}, \bibinfo{eid}{046003}
  (\bibinfo{year}{2010}), \eprint{0808.0706}.

\bibitem[{\citenamefont{{Flauger} et~al.}(2010)\citenamefont{{Flauger},
  {McAllister}, {Pajer}, {Westphal}, and {Xu}}}]{Flauger2010}
\bibinfo{author}{\bibfnamefont{R.}~\bibnamefont{{Flauger}}},
  \bibinfo{author}{\bibfnamefont{L.}~\bibnamefont{{McAllister}}},
  \bibinfo{author}{\bibfnamefont{E.}~\bibnamefont{{Pajer}}},
  \bibinfo{author}{\bibfnamefont{A.}~\bibnamefont{{Westphal}}},
  \bibnamefont{and} \bibinfo{author}{\bibfnamefont{G.}~\bibnamefont{{Xu}}},
  \bibinfo{journal}{\jcap} \textbf{\bibinfo{volume}{6}}, \bibinfo{eid}{009}
  (\bibinfo{year}{2010}), \eprint{0907.2916}.

\bibitem[{\citenamefont{Freese et~al.}(1990)\citenamefont{Freese, Frieman, and
  Olinto}}]{PhysRevLett.65.3233}
\bibinfo{author}{\bibfnamefont{K.}~\bibnamefont{Freese}},
  \bibinfo{author}{\bibfnamefont{J.~A.} \bibnamefont{Frieman}},
  \bibnamefont{and} \bibinfo{author}{\bibfnamefont{A.~V.}
  \bibnamefont{Olinto}}, \bibinfo{journal}{Phys. Rev. Lett.}
  \textbf{\bibinfo{volume}{65}}, \bibinfo{pages}{3233} (\bibinfo{year}{1990}),
  \urlprefix\url{http://link.aps.org/doi/10.1103/PhysRevLett.65.3233}.

\bibitem[{\citenamefont{{Meerburg} et~al.}(2012)\citenamefont{{Meerburg},
  {Wijers}, and {van der Schaar}}}]{Meerburg2012}
\bibinfo{author}{\bibfnamefont{P.~D.} \bibnamefont{{Meerburg}}},
  \bibinfo{author}{\bibfnamefont{R.~A.~M.~J.} \bibnamefont{{Wijers}}},
  \bibnamefont{and} \bibinfo{author}{\bibfnamefont{J.~P.} \bibnamefont{{van der
  Schaar}}}, \bibinfo{journal}{\mnras} \textbf{\bibinfo{volume}{421}},
  \bibinfo{pages}{369} (\bibinfo{year}{2012}), \eprint{1109.5264}.

\bibitem[{\citenamefont{{Easther} and {Flauger}}(2013)}]{2013Flaugera}
\bibinfo{author}{\bibfnamefont{R.}~\bibnamefont{{Easther}}} \bibnamefont{and}
  \bibinfo{author}{\bibfnamefont{R.}~\bibnamefont{{Flauger}}},
  \bibinfo{journal}{ArXiv e-prints}  (\bibinfo{year}{2013}),
  \eprint{1308.3736}.

\bibitem[{\citenamefont{{Meerburg}
  et~al.}(2014{\natexlab{a}})\citenamefont{{Meerburg}, {Spergel}, and
  {Wandelt}}}]{Meerburg2014b}
\bibinfo{author}{\bibfnamefont{P.~D.} \bibnamefont{{Meerburg}}},
  \bibinfo{author}{\bibfnamefont{D.~N.} \bibnamefont{{Spergel}}},
  \bibnamefont{and} \bibinfo{author}{\bibfnamefont{B.~D.}
  \bibnamefont{{Wandelt}}}, \bibinfo{journal}{\prd}
  \textbf{\bibinfo{volume}{89}}, \bibinfo{eid}{063536}
  (\bibinfo{year}{2014}{\natexlab{a}}), \eprint{1308.3704}.

\bibitem[{\citenamefont{{Meerburg}
  et~al.}(2014{\natexlab{b}})\citenamefont{{Meerburg}, {Spergel}, and
  {Wandelt}}}]{Meerburg2014a}
\bibinfo{author}{\bibfnamefont{P.~D.} \bibnamefont{{Meerburg}}},
  \bibinfo{author}{\bibfnamefont{D.~N.} \bibnamefont{{Spergel}}},
  \bibnamefont{and} \bibinfo{author}{\bibfnamefont{B.~D.}
  \bibnamefont{{Wandelt}}}, \bibinfo{journal}{\prd}
  \textbf{\bibinfo{volume}{89}}, \bibinfo{eid}{063537}
  (\bibinfo{year}{2014}{\natexlab{b}}), \eprint{1308.3705}.

\bibitem[{\citenamefont{{Meerburg}
  et~al.}(2014{\natexlab{c}})\citenamefont{{Meerburg}, {Spergel}, and
  {Wandelt}}}]{Meerburg2014c}
\bibinfo{author}{\bibfnamefont{P.~D.} \bibnamefont{{Meerburg}}},
  \bibinfo{author}{\bibfnamefont{D.~N.} \bibnamefont{{Spergel}}},
  \bibnamefont{and} \bibinfo{author}{\bibfnamefont{B.~D.}
  \bibnamefont{{Wandelt}}}, \bibinfo{journal}{ArXiv e-prints}
  (\bibinfo{year}{2014}{\natexlab{c}}), \eprint{1406.0548}.

\bibitem[{\citenamefont{{Peiris} et~al.}(2013)\citenamefont{{Peiris},
  {Easther}, and {Flauger}}}]{Flauger2013}
\bibinfo{author}{\bibfnamefont{H.}~\bibnamefont{{Peiris}}},
  \bibinfo{author}{\bibfnamefont{R.}~\bibnamefont{{Easther}}},
  \bibnamefont{and}
  \bibinfo{author}{\bibfnamefont{R.}~\bibnamefont{{Flauger}}},
  \bibinfo{journal}{ArXiv e-prints}  (\bibinfo{year}{2013}),
  \eprint{1303.2616}.

\bibitem[{\citenamefont{{Pajer} and {Peloso}}(2013)}]{PajerPeloso2013}
\bibinfo{author}{\bibfnamefont{E.}~\bibnamefont{{Pajer}}} \bibnamefont{and}
  \bibinfo{author}{\bibfnamefont{M.}~\bibnamefont{{Peloso}}},
  \bibinfo{journal}{Classical and Quantum Gravity}
  \textbf{\bibinfo{volume}{30}}, \bibinfo{eid}{214002} (\bibinfo{year}{2013}),
  \eprint{1305.3557}.

\bibitem[{\citenamefont{{Baumann} and {McAllister}}(2014)}]{DanLiam2014}
\bibinfo{author}{\bibfnamefont{D.}~\bibnamefont{{Baumann}}} \bibnamefont{and}
  \bibinfo{author}{\bibfnamefont{L.}~\bibnamefont{{McAllister}}},
  \bibinfo{journal}{ArXiv e-prints}  (\bibinfo{year}{2014}),
  \eprint{1404.2601}.

\bibitem[{\citenamefont{{Greene} et~al.}(2005)\citenamefont{{Greene}, {Schalm},
  {van der Schaar}, and {Shiu}}}]{Greene2005}
\bibinfo{author}{\bibfnamefont{B.}~\bibnamefont{{Greene}}},
  \bibinfo{author}{\bibfnamefont{K.}~\bibnamefont{{Schalm}}},
  \bibinfo{author}{\bibfnamefont{J.~P.} \bibnamefont{{van der Schaar}}},
  \bibnamefont{and} \bibinfo{author}{\bibfnamefont{G.}~\bibnamefont{{Shiu}}},
  in \emph{\bibinfo{booktitle}{22nd Texas Symposium on Relativistic
  Astrophysics}}, edited by
  \bibinfo{editor}{\bibfnamefont{P.}~\bibnamefont{{Chen}}},
  \bibinfo{editor}{\bibfnamefont{E.}~\bibnamefont{{Bloom}}},
  \bibinfo{editor}{\bibfnamefont{G.}~\bibnamefont{{Madejski}}},
  \bibnamefont{and}
  \bibinfo{editor}{\bibfnamefont{V.}~\bibnamefont{{Patrosian}}}
  (\bibinfo{year}{2005}), pp. \bibinfo{pages}{1--8},
  \eprint{arXiv:astro-ph/0503458}.

\bibitem[{\citenamefont{{D'Amico} et~al.}(2013)\citenamefont{{D'Amico},
  {Gobbetti}, {Kleban}, and {Schillo}}}]{Damico2013}
\bibinfo{author}{\bibfnamefont{G.}~\bibnamefont{{D'Amico}}},
  \bibinfo{author}{\bibfnamefont{R.}~\bibnamefont{{Gobbetti}}},
  \bibinfo{author}{\bibfnamefont{M.}~\bibnamefont{{Kleban}}}, \bibnamefont{and}
  \bibinfo{author}{\bibfnamefont{M.}~\bibnamefont{{Schillo}}},
  \bibinfo{journal}{\jcap} \textbf{\bibinfo{volume}{3}}, \bibinfo{eid}{004}
  (\bibinfo{year}{2013}), \eprint{1211.4589}.

\bibitem[{\citenamefont{{Ashoorioon} et~al.}(2009)\citenamefont{{Ashoorioon},
  {Krause}, and {Turzynski}}}]{Ashoorioon2009}
\bibinfo{author}{\bibfnamefont{A.}~\bibnamefont{{Ashoorioon}}},
  \bibinfo{author}{\bibfnamefont{A.}~\bibnamefont{{Krause}}}, \bibnamefont{and}
  \bibinfo{author}{\bibfnamefont{K.}~\bibnamefont{{Turzynski}}},
  \bibinfo{journal}{\jcap} \textbf{\bibinfo{volume}{2}}, \bibinfo{eid}{014}
  (\bibinfo{year}{2009}), \eprint{0810.4660}.

\bibitem[{\citenamefont{{Ach{\'u}carro}
  et~al.}(2014)\citenamefont{{Ach{\'u}carro}, {Atal}, {Ortiz}, and
  {Torrado}}}]{2014PhRvD..89j3006A}
\bibinfo{author}{\bibfnamefont{A.}~\bibnamefont{{Ach{\'u}carro}}},
  \bibinfo{author}{\bibfnamefont{V.}~\bibnamefont{{Atal}}},
  \bibinfo{author}{\bibfnamefont{P.}~\bibnamefont{{Ortiz}}}, \bibnamefont{and}
  \bibinfo{author}{\bibfnamefont{J.}~\bibnamefont{{Torrado}}},
  \bibinfo{journal}{\prd} \textbf{\bibinfo{volume}{89}}, \bibinfo{eid}{103006}
  (\bibinfo{year}{2014}), \eprint{1311.2552}.

\bibitem[{\citenamefont{{Hazra} et~al.}(2013)\citenamefont{{Hazra},
  {Shafieloo}, and {Smoot}}}]{2013JCAP...12..035H}
\bibinfo{author}{\bibfnamefont{D.~K.} \bibnamefont{{Hazra}}},
  \bibinfo{author}{\bibfnamefont{A.}~\bibnamefont{{Shafieloo}}},
  \bibnamefont{and} \bibinfo{author}{\bibfnamefont{G.~F.}
  \bibnamefont{{Smoot}}}, \bibinfo{journal}{\jcap}
  \textbf{\bibinfo{volume}{12}}, \bibinfo{eid}{035} (\bibinfo{year}{2013}),
  \eprint{1310.3038}.

\bibitem[{\citenamefont{{Planck collaboration}
  et~al.}(2013)\citenamefont{{Planck collaboration}, {Ade}, {Aghanim},
  {Armitage-Caplan}, {Arnaud}, {Ashdown}, {Atrio-Barandela}, {Aumont},
  {Baccigalupi}, {Banday} et~al.}}]{planckLikelihood2013}
\bibinfo{author}{\bibnamefont{{Planck collaboration}}},
  \bibinfo{author}{\bibfnamefont{P.~A.~R.} \bibnamefont{{Ade}}},
  \bibinfo{author}{\bibfnamefont{N.}~\bibnamefont{{Aghanim}}},
  \bibinfo{author}{\bibfnamefont{C.}~\bibnamefont{{Armitage-Caplan}}},
  \bibinfo{author}{\bibfnamefont{M.}~\bibnamefont{{Arnaud}}},
  \bibinfo{author}{\bibfnamefont{M.}~\bibnamefont{{Ashdown}}},
  \bibinfo{author}{\bibfnamefont{F.}~\bibnamefont{{Atrio-Barandela}}},
  \bibinfo{author}{\bibfnamefont{J.}~\bibnamefont{{Aumont}}},
  \bibinfo{author}{\bibfnamefont{C.}~\bibnamefont{{Baccigalupi}}},
  \bibinfo{author}{\bibfnamefont{A.~J.} \bibnamefont{{Banday}}},
  \bibnamefont{et~al.}, \bibinfo{journal}{ArXiv e-prints}
  (\bibinfo{year}{2013}), \eprint{1303.5075}.

\bibitem[{\citenamefont{{Lewis} and {Bridle}}(2002)}]{Lewis2002}
\bibinfo{author}{\bibfnamefont{A.}~\bibnamefont{{Lewis}}} \bibnamefont{and}
  \bibinfo{author}{\bibfnamefont{S.}~\bibnamefont{{Bridle}}},
  \bibinfo{journal}{\prd} \textbf{\bibinfo{volume}{66}}, \bibinfo{eid}{103511}
  (\bibinfo{year}{2002}), \eprint{arXiv:astro-ph/0205436}.

\bibitem[{\citenamefont{{Das} et~al.}(2014)\citenamefont{{Das}, {Louis},
  {Nolta}, {Addison}, {Battistelli}, {Bond}, {Calabrese}, {Crichton}, {Devlin},
  {Dicker} et~al.}}]{ACT2014}
\bibinfo{author}{\bibfnamefont{S.}~\bibnamefont{{Das}}},
  \bibinfo{author}{\bibfnamefont{T.}~\bibnamefont{{Louis}}},
  \bibinfo{author}{\bibfnamefont{M.~R.} \bibnamefont{{Nolta}}},
  \bibinfo{author}{\bibfnamefont{G.~E.} \bibnamefont{{Addison}}},
  \bibinfo{author}{\bibfnamefont{E.~S.} \bibnamefont{{Battistelli}}},
  \bibinfo{author}{\bibfnamefont{J.~R.} \bibnamefont{{Bond}}},
  \bibinfo{author}{\bibfnamefont{E.}~\bibnamefont{{Calabrese}}},
  \bibinfo{author}{\bibfnamefont{D.}~\bibnamefont{{Crichton}}},
  \bibinfo{author}{\bibfnamefont{M.~J.} \bibnamefont{{Devlin}}},
  \bibinfo{author}{\bibfnamefont{S.}~\bibnamefont{{Dicker}}},
  \bibnamefont{et~al.}, \bibinfo{journal}{\jcap} \textbf{\bibinfo{volume}{4}},
  \bibinfo{eid}{014} (\bibinfo{year}{2014}), \eprint{1301.1037}.

\bibitem[{\citenamefont{{Keisler} et~al.}(2011)\citenamefont{{Keisler},
  {Reichardt}, {Aird}, {Benson}, {Bleem}, {Carlstrom}, {Chang}, {Cho},
  {Crawford}, {Crites} et~al.}}]{SPT2011}
\bibinfo{author}{\bibfnamefont{R.}~\bibnamefont{{Keisler}}},
  \bibinfo{author}{\bibfnamefont{C.~L.} \bibnamefont{{Reichardt}}},
  \bibinfo{author}{\bibfnamefont{K.~A.} \bibnamefont{{Aird}}},
  \bibinfo{author}{\bibfnamefont{B.~A.} \bibnamefont{{Benson}}},
  \bibinfo{author}{\bibfnamefont{L.~E.} \bibnamefont{{Bleem}}},
  \bibinfo{author}{\bibfnamefont{J.~E.} \bibnamefont{{Carlstrom}}},
  \bibinfo{author}{\bibfnamefont{C.~L.} \bibnamefont{{Chang}}},
  \bibinfo{author}{\bibfnamefont{H.~M.} \bibnamefont{{Cho}}},
  \bibinfo{author}{\bibfnamefont{T.~M.} \bibnamefont{{Crawford}}},
  \bibinfo{author}{\bibfnamefont{A.~T.} \bibnamefont{{Crites}}},
  \bibnamefont{et~al.}, \bibinfo{journal}{\apj} \textbf{\bibinfo{volume}{743}},
  \bibinfo{eid}{28} (\bibinfo{year}{2011}), \eprint{1105.3182}.

\bibitem[{\citenamefont{{Reichardt} et~al.}(2012)\citenamefont{{Reichardt},
  {Shaw}, {Zahn}, {Aird}, {Benson}, {Bleem}, {Carlstrom}, {Chang}, {Cho},
  {Crawford} et~al.}}]{SPT2012}
\bibinfo{author}{\bibfnamefont{C.~L.} \bibnamefont{{Reichardt}}},
  \bibinfo{author}{\bibfnamefont{L.}~\bibnamefont{{Shaw}}},
  \bibinfo{author}{\bibfnamefont{O.}~\bibnamefont{{Zahn}}},
  \bibinfo{author}{\bibfnamefont{K.~A.} \bibnamefont{{Aird}}},
  \bibinfo{author}{\bibfnamefont{B.~A.} \bibnamefont{{Benson}}},
  \bibinfo{author}{\bibfnamefont{L.~E.} \bibnamefont{{Bleem}}},
  \bibinfo{author}{\bibfnamefont{J.~E.} \bibnamefont{{Carlstrom}}},
  \bibinfo{author}{\bibfnamefont{C.~L.} \bibnamefont{{Chang}}},
  \bibinfo{author}{\bibfnamefont{H.~M.} \bibnamefont{{Cho}}},
  \bibinfo{author}{\bibfnamefont{T.~M.} \bibnamefont{{Crawford}}},
  \bibnamefont{et~al.}, \bibinfo{journal}{\apj} \textbf{\bibinfo{volume}{755}},
  \bibinfo{eid}{70} (\bibinfo{year}{2012}), \eprint{1111.0932}.

\bibitem[{\citenamefont{{Feroz} et~al.}(2009)\citenamefont{{Feroz}, {Hobson},
  and {Bridges}}}]{Feroz2009}
\bibinfo{author}{\bibfnamefont{F.}~\bibnamefont{{Feroz}}},
  \bibinfo{author}{\bibfnamefont{M.~P.} \bibnamefont{{Hobson}}},
  \bibnamefont{and}
  \bibinfo{author}{\bibfnamefont{M.}~\bibnamefont{{Bridges}}},
  \bibinfo{journal}{\mnras} \textbf{\bibinfo{volume}{398}},
  \bibinfo{pages}{1601} (\bibinfo{year}{2009}), \eprint{0809.3437}.

\bibitem[{\citenamefont{{Feroz} et~al.}(2013)\citenamefont{{Feroz}, {Hobson},
  {Cameron}, and {Pettitt}}}]{Feroz2013}
\bibinfo{author}{\bibfnamefont{F.}~\bibnamefont{{Feroz}}},
  \bibinfo{author}{\bibfnamefont{M.~P.} \bibnamefont{{Hobson}}},
  \bibinfo{author}{\bibfnamefont{E.}~\bibnamefont{{Cameron}}},
  \bibnamefont{and} \bibinfo{author}{\bibfnamefont{A.~N.}
  \bibnamefont{{Pettitt}}}, \bibinfo{journal}{ArXiv e-prints}
  (\bibinfo{year}{2013}), \eprint{1306.2144}.

\bibitem[{\citenamefont{{Spergel} et~al.}(2013)\citenamefont{{Spergel},
  {Gehrels}, {Breckinridge}, {Donahue}, {Dressler}, {Gaudi}, {Greene}, {Guyon},
  {Hirata}, {Kalirai} et~al.}}]{WFIRST2013}
\bibinfo{author}{\bibfnamefont{D.}~\bibnamefont{{Spergel}}},
  \bibinfo{author}{\bibfnamefont{N.}~\bibnamefont{{Gehrels}}},
  \bibinfo{author}{\bibfnamefont{J.}~\bibnamefont{{Breckinridge}}},
  \bibinfo{author}{\bibfnamefont{M.}~\bibnamefont{{Donahue}}},
  \bibinfo{author}{\bibfnamefont{A.}~\bibnamefont{{Dressler}}},
  \bibinfo{author}{\bibfnamefont{B.~S.} \bibnamefont{{Gaudi}}},
  \bibinfo{author}{\bibfnamefont{T.}~\bibnamefont{{Greene}}},
  \bibinfo{author}{\bibfnamefont{O.}~\bibnamefont{{Guyon}}},
  \bibinfo{author}{\bibfnamefont{C.}~\bibnamefont{{Hirata}}},
  \bibinfo{author}{\bibfnamefont{J.}~\bibnamefont{{Kalirai}}},
  \bibnamefont{et~al.}, \bibinfo{journal}{ArXiv e-prints}
  (\bibinfo{year}{2013}), \eprint{1305.5422}.

\bibitem[{\citenamefont{{McAllister} et~al.}(2014)\citenamefont{{McAllister},
  {Silverstein}, {Westphal}, and {Wrase}}}]{2014arXiv1405.3652M}
\bibinfo{author}{\bibfnamefont{L.}~\bibnamefont{{McAllister}}},
  \bibinfo{author}{\bibfnamefont{E.}~\bibnamefont{{Silverstein}}},
  \bibinfo{author}{\bibfnamefont{A.}~\bibnamefont{{Westphal}}},
  \bibnamefont{and} \bibinfo{author}{\bibfnamefont{T.}~\bibnamefont{{Wrase}}},
  \bibinfo{journal}{ArXiv e-prints}  (\bibinfo{year}{2014}),
  \eprint{1405.3652}.

\bibitem[{\citenamefont{Weinberg}(2008)}]{steven2008cosmology}
\bibinfo{author}{\bibfnamefont{S.}~\bibnamefont{Weinberg}},
  \emph{\bibinfo{title}{Cosmology}}, Cosmology (\bibinfo{publisher}{OUP
  Oxford}, \bibinfo{year}{2008}), ISBN \bibinfo{isbn}{9780191523601},
  \urlprefix\url{http://books.google.com/books?id=nqQZdg020fsC}.

\bibitem[{\citenamefont{{Flauger} and {Pajer}}(2011)}]{FlaugerPajer2011}
\bibinfo{author}{\bibfnamefont{R.}~\bibnamefont{{Flauger}}} \bibnamefont{and}
  \bibinfo{author}{\bibfnamefont{E.}~\bibnamefont{{Pajer}}},
  \bibinfo{journal}{\jcap} \textbf{\bibinfo{volume}{1}}, \bibinfo{eid}{017}
  (\bibinfo{year}{2011}), \eprint{1002.0833}.

\bibitem[{\citenamefont{{Audren} et~al.}(2014)\citenamefont{{Audren},
  {Figueroa}, and {Tram}}}]{Audren2014}
\bibinfo{author}{\bibfnamefont{B.}~\bibnamefont{{Audren}}},
  \bibinfo{author}{\bibfnamefont{D.~G.} \bibnamefont{{Figueroa}}},
  \bibnamefont{and} \bibinfo{author}{\bibfnamefont{T.}~\bibnamefont{{Tram}}},
  \bibinfo{journal}{ArXiv e-prints}  (\bibinfo{year}{2014}),
  \eprint{1405.1390}.

\bibitem[{\citenamefont{{Choudhury} and {Mazumdar}}(2014)}]{Choudhury2014}
\bibinfo{author}{\bibfnamefont{S.}~\bibnamefont{{Choudhury}}} \bibnamefont{and}
  \bibinfo{author}{\bibfnamefont{A.}~\bibnamefont{{Mazumdar}}},
  \bibinfo{journal}{ArXiv e-prints}  (\bibinfo{year}{2014}),
  \eprint{1403.5549}.

\bibitem[{\citenamefont{{Wan} et~al.}(2014)\citenamefont{{Wan}, {Li}, {Li},
  {Qiu}, {Cai}, and {Zhang}}}]{2014arXiv1405.2784W}
\bibinfo{author}{\bibfnamefont{Y.}~\bibnamefont{{Wan}}},
  \bibinfo{author}{\bibfnamefont{S.}~\bibnamefont{{Li}}},
  \bibinfo{author}{\bibfnamefont{M.}~\bibnamefont{{Li}}},
  \bibinfo{author}{\bibfnamefont{T.}~\bibnamefont{{Qiu}}},
  \bibinfo{author}{\bibfnamefont{Y.}~\bibnamefont{{Cai}}}, \bibnamefont{and}
  \bibinfo{author}{\bibfnamefont{X.}~\bibnamefont{{Zhang}}},
  \bibinfo{journal}{ArXiv e-prints}  (\bibinfo{year}{2014}),
  \eprint{1405.2784}.

\bibitem[{\citenamefont{{Freese} and {Kinney}}(2014)}]{2014arXiv1403.5277F}
\bibinfo{author}{\bibfnamefont{K.}~\bibnamefont{{Freese}}} \bibnamefont{and}
  \bibinfo{author}{\bibfnamefont{W.~H.} \bibnamefont{{Kinney}}},
  \bibinfo{journal}{ArXiv e-prints}  (\bibinfo{year}{2014}),
  \eprint{1403.5277}.

\bibitem[{\citenamefont{{Barnaby} et~al.}(2011)\citenamefont{{Barnaby},
  {Namba}, and {Peloso}}}]{2011JCAP...04..009B}
\bibinfo{author}{\bibfnamefont{N.}~\bibnamefont{{Barnaby}}},
  \bibinfo{author}{\bibfnamefont{R.}~\bibnamefont{{Namba}}}, \bibnamefont{and}
  \bibinfo{author}{\bibfnamefont{M.}~\bibnamefont{{Peloso}}},
  \bibinfo{journal}{\jcap} \textbf{\bibinfo{volume}{4}}, \bibinfo{eid}{009}
  (\bibinfo{year}{2011}), \eprint{1102.4333}.

\bibitem[{\citenamefont{{Meerburg} and {Pajer}}(2013)}]{2013JCAP...02..017M}
\bibinfo{author}{\bibfnamefont{P.~D.} \bibnamefont{{Meerburg}}}
  \bibnamefont{and} \bibinfo{author}{\bibfnamefont{E.}~\bibnamefont{{Pajer}}},
  \bibinfo{journal}{\jcap} \textbf{\bibinfo{volume}{2}}, \bibinfo{eid}{017}
  (\bibinfo{year}{2013}), \eprint{1203.6076}.

\bibitem[{\citenamefont{{Linde} et~al.}(2013)\citenamefont{{Linde}, {Mooij},
  and {Pajer}}}]{2013PhRvD..87j3506L}
\bibinfo{author}{\bibfnamefont{A.}~\bibnamefont{{Linde}}},
  \bibinfo{author}{\bibfnamefont{S.}~\bibnamefont{{Mooij}}}, \bibnamefont{and}
  \bibinfo{author}{\bibfnamefont{E.}~\bibnamefont{{Pajer}}},
  \bibinfo{journal}{\prd} \textbf{\bibinfo{volume}{87}}, \bibinfo{eid}{103506}
  (\bibinfo{year}{2013}), \eprint{1212.1693}.

\end{thebibliography}

\end{document}